\documentclass[onecolumn,superscriptaddress,prc]{revtex4}
\usepackage{graphicx}
\usepackage{diagbox}
\newcommand{\pt}{p_{\rm T}}
\usepackage{bm}

\begin{document}
\title{Validation and improvement of the ZPC parton cascade inside a box}
\author{Xin-Li Zhao}
\affiliation{Key Laboratory of Nuclear Physics and Ion-beam
  Application (MOE), Institute of Modern Physics, Fudan University,
  Shanghai 200433, China}
\affiliation{Department of Physics, East Carolina University,
  Greenville, North Carolina 27858, USA} 
\affiliation{Shanghai Institute of Applied Physics, Chinese Academy of
  Sciences, Shanghai 201800, China} 
\affiliation{University of Chinese Academy of Sciences, Beijing 100049, China}
\author{Guo-Liang Ma}
\affiliation{Key Laboratory of Nuclear Physics and Ion-beam
  Application (MOE), Institute of Modern Physics, Fudan University,
  Shanghai 200433, China}
\author{Yu-Gang Ma}
\affiliation{Key Laboratory of Nuclear Physics and Ion-beam
  Application (MOE), Institute of Modern Physics, Fudan University,
  Shanghai 200433, China} 
\author{Zi-Wei Lin}
\email[]{linz@ecu.edu}
\affiliation{Department of Physics, East Carolina University,
  Greenville, North Carolina 27858, USA} 

\begin{abstract}
Cascade solutions of the Boltzmann equation suffer from causality
violation at large densities and/or scattering cross sections. Although
the particle subdivision technique can reduce the causality violation,
it alters event-by-event correlations and fluctuations and is also
computationally expensive.
Here we evaluate and then improve the accuracy of the ZPC parton
cascade for elastic scatterings inside a box without using parton
subdivision. We first test different collision schemes for the
collision times and ordering time and find that the default
collision scheme does not accurately describe the equilibrium momentum
distribution at large opacities. We then find a specific collision
scheme that can describe very accurately the equilibrium momentum 
distribution as well as the time evolution towards equilibrium, even
at large opacities. We also calculate the shear viscosity and the
$\eta/s$ ratio of the parton systems and confirm that the new collision
scheme is more accurate. 
In addition, we use a novel parton subdivision
method to obtain the ``exact'' evolution of the system. This
subdivision method is valid for such box calculations and is so much 
more efficient than the standard subdivision method that we use a
subdivision factor of $10^6$ in this study.
\end{abstract}

\maketitle

\section{Introduction}
\label{introduction}

In high energy heavy ion collisions such as those at the Relativistic
Heavy Ion Collider (RHIC) and the Large Hadron Collider (LHC), the
quark-gluon plasma (QGP) with deconfined parton degrees of freedom is
formed~\cite{Adams:2005dq,Adcox:2004mh}.
Interactions among the partons, which reflect the properties of the
QGP, could significantly affect many final state observables such as
the hadron spectra, collective flows, and fluctuations
\cite{He:2015hfa,Lin:2015ucn,Ko:2016ioz,Jin:2018lbk,Wang:2019vhg}. 
A parton cascade
model provides a microscopic description of the space-time evolution
of the partonic phase of relativistic heavy ion collisions. 
Both elastic and inelastic parton cascade models, such as
VNI~\cite{Geiger:1997pf}, ZPC~\cite{Gyulassy:1997zn,Zhang:1997ej},
MPC~\cite{Molnar:2000jh}, and BAMPS~\cite{Xu:2004mz,Xu:2007aa}, have
been constructed to model parton interactions. 
For example, recent studies from a multi-phase transport (AMPT)
model~\cite{He:2015hfa,Lin:2015ucn}, which includes the ZPC elastic
parton cascade~\cite{Lin:2004en}, have shown that even a few parton
scatterings in a small system is enough to generate significant
momentum anisotropies~\cite{Bzdak:2014dia,Li:2016ubw}.
This concerns the origin of collectivity and the difference between
kinetic theory and hydrodynamics in heavy ion collisions, particularly
in small systems~\cite{Nagle:2017sjv,Kurkela3}.
It is therefore important to ensure that the parton cascade solution is accurate in solving the corresponding Boltzmann equation.

The ZPC elastic parton cascade~\cite{Zhang:1997ej,Gyulassy:1997zn} solves the Boltzmann equation by the cascade method. A scattering happens when the closest distance between two partons is less than the range of interaction $\sqrt{\sigma/\pi}$, where $\sigma$ is the parton scattering cross section.
It is well known that causality
violation~\cite{Kodama:1983yk,Kortemeyer:1995di} is inherent in
cascade simulations due to the geometrical interpretation of cross
section. This leads to inaccurate numerical results at large opacities,
i.e., at high densities and/or large scattering cross sections.
For example, a recent study~\cite{Molnar:2019yam} has shown that the
effect of causality violation on the elliptic flow from the string
melting version of the AMPT model~\cite{Lin:2004en} is small but
non-zero. This is mainly because the parton density is very
high~\cite{Lin:2014tya} even though the cross section is small ($\sim
3$ mb).  Causality violation also leads to the fact that different
choices of doing collisions and/or the reference frame can lead to
different numerical
results~\cite{Zhang:1996gb,Zhang:1998tj,Cheng:2001dz}.
These numerical artifacts due to the causality violation can be
reduced or removed by the parton subdivision
technique (i.e., the test particle multiplication
method)~\cite{Wong:1982zzb,Welke:1989dr,Kortemeyer:1995di,Pang:1997,Zhang:1998tj,Molnar:2000jh,Molnar:2001ux,Molnar:2004yh,Xu:2004mz}. 
However, parton subdivision alters the event-by-event correlations and
fluctuations, the importance of which has been more appreciated in
recent years~\cite{Alver:2010gr}; parton subdivision is also much more
computationally expensive.

Therefore the goal of this work is to find a parton cascade
algorithm that is accurate enough without using parton subdivision.
We investigate different collision schemes for the ZPC parton cascade
for elastic scatterings in a box with periodic boundaries 
and then compare the results with
either the theoretical expectation or the ``exact'' results from ZPC
with parton subdivision. The paper is organized as follows.
In Sec.~\ref{zpc} we give a brief introduction to the ZPC parton
cascade. We then discuss the parton subdivision technique in
Sec.~\ref{subd}. Numerical results of the $\pt$ distribution and the
shear viscosity including the $\eta/s$ ratio for several cases 
are presented and discussed in Sec.~\ref{results}. Finally we conclude
in Sec.~\ref{summary}. 

\section{The ZPC parton cascade}
\label{zpc}

The ZPC parton cascade~\cite{Zhang:1997ej,Gyulassy:1997zn} includes two-body elastic parton scatterings such as $gg\rightarrow gg$ by solving the Boltzmann equation, where the on-shell phase space density $f(\bm r,\bm p, t)$ evolves as
\begin{equation}
p^\mu \partial_\mu f (\bm r, \bm p, t)
= \mathcal C \left [ |\mathcal M|^2 f_1(\bm r_1, \bm p_1, t) f_2(\bm
  r_2, \bm p_2, t) \right ].
\label{be1}
\end{equation}
In the above, the collision term $\mathcal C[~]$  includes the integral over
the momenta of the other three partons with an integrand containing
factors such as a $\delta$-function for the energy-momentum
conservation. The differential cross section of parton scatterings is
given by the matrix element as $d\sigma/d \hat t \propto  |\mathcal M|^2$.

The default differential cross section in ZPC for two-parton
scatterings, based on the gluon elastic scattering cross section as
calculated by leading-order QCD, is given
by~\cite{Zhang:1997ej,Lin:2004en}
\begin{equation}
\frac{d\sigma}{d\hat t}=\frac{9\pi \alpha_{s}^{2}}{2} \left
  (1+\frac{\mu^2}{\hat s} \right ) \frac{1}{(\hat t-\mu ^{2})^{2}},
\label{dsigmadt}
\end{equation}
where $\alpha_{s}$ is the strong coupling constant, $\hat s$ and $\hat
t$ are the standard Mandelstam variables, and $\mu$ is a screening mass to regular the total cross section.
This way the total cross section has no explicit dependence on $\hat s$ as
\begin{equation}
\sigma = \frac{9\pi \alpha_{s}^{2}}{2\mu ^{2}}.
\label{sigma}
\end{equation}
The above Eq.(\ref{dsigmadt}) represents forward-angle scatterings. We
also test isotropic scatterings in this study, where $d\sigma/d\hat t$ is
independent of the scattering angle.
For this study we take $\alpha_{s}=\sqrt
{2/9}$~\cite{Zhang:1997ej} unless specified otherwise.

In ZPC one can take different choices or collision schemes to
implement the cascade method~\cite{Zhang:1997ej}, and ZPC already
provides several different choices.
With the closest approach criterion for parton scatterings, the
closest approach distance may be calculated either in the two-parton
center of mass frame or in the global frame of the whole parton
system of each event.
Two partons may collide when their closest approach distance is
smaller than $\sqrt{\sigma/\pi}$, and at a given global time all such
possible collisions in the future are ordered in a collision list with
the ordering time of each collision, so that
they can be carried out sequentially. The collision list is updated
continuously after each collision,
and for expansion cases the parton system dynamically freezes out when
the collision list is empty.
For box cases in this work, we terminate the parton cascade at a
global time that is large enough so that the parton momentum
distribution changes little afterwards.
When the closest approach distance is calculated in the two-parton
center of mass frame, the collision time of a scattering in that frame
is a well-defined single value.
However, because of the finite $\sigma$ the two partons have different
spatial coordinates in general, therefore this collision time in the
two-parton center of mass frame becomes two different colliding times
in the global frame (named here as $ct_1$ and $ct_2$ respectively for
the two colliding partons) after the Lorentz transformation. 
Note that each of the two partons involved in a scattering
changes its momentum at its collision time at the corresponding
position in the global frame.

We show in Table~\ref{schemes} a dozen different collision schemes
for the case of calculating the closest approach distance in the
two-parton center of mass frame, where a collision scheme refers to a
given choice of the collision time(s) and the ordering time. These
schemes include the ones that choose the collision time(s) in the global
frame as separate values (i.e., as $ct_1$ and $ct_2$), the earlier time
$min(ct_1,ct_2)$, the average time $(ct_1+ct_2)/2$, or the later time
$max(ct_1,ct_2)$, in combination with choosing the collision ordering
time  in the global frame as either the earlier time, the average
time, or the later time.
On the other hand, for the case of calculating the closest
approach distance in the global frame, it is natural to choose the
single collision time as the collision ordering time (both in the
global frame); this is called collision scheme M here.

\begin{table}[h]
\caption{Different collision schemes for ZPC when the closest approach distance is calculated in the two-parton center of mass frame.
$ct_1$ and $ct_2$ represent the collision times of the two scattered
partons after the transformation back to the global frame.}
\renewcommand\arraystretch{1.5}
\begin{tabular}{|c|c|c|c|c|}
\hline
\diagbox{Ordering time}{Collision time} & $ct_1$ \& $ct_2$  & $min(ct_1,ct_2)$  & $(ct_1+ct_2)/2$ & $max(ct_1,ct_2)$\\
\hline
$min(ct_1,ct_2)$  & A   & B (new scheme)  & C   & D  \\
\hline
$(ct_1+ct_2)/2$ & E & F  & G (default ZPC scheme) & H \\
\hline
$max(ct_1,ct_2)$ & I  & J   & K  & L \\
\hline
\end{tabular}
\label{schemes}
\end{table}

We first test these different collision schemes for the case of 4,000
massless gluons in a box with a cross section of $\sigma=2.6$ mb for
forward-angle scatterings. The box size is chosen such that the gluon density
$n$ is the same as that for a thermalized gluon system at the
temperature $T=0.5$ GeV, i.e., $n=d_gT^3/\pi^2$ with the gluon 
degeneracy factor $d_g=16$. 
Since we do not include quantum statistics
in the collision kernel of Eq.(\ref{be1}), the final state momentum
distribution of the gluon system should be given by the
Maxwell$\textendash$Boltzmann distribution at this temperature.

We use an off-equilibrium initial condition
that is uniform in the coordinate space, which is obtained by
first generating the Maxwell$\textendash$Boltzmann momentum
distribution and then decreasing each parton's initial $p_z$ by a
factor of two while keeping its initial 3-momentum the same.
Such off-equilibrium initial momentum distribution is used for all the
calculations of the $\pt$ spectra in Sec.~\ref{resultsA}, 
while the equilibrium initial momentum distribution is 
used for calculations of shear viscosity in Sec.~\ref{resultsB}. 
Also note that the ZPC results in this
study are obtained without using parton subdivision unless specified
otherwise, and typically a few thousand events is used for each
case while each event simulates the scatterings of at least 4,000
massless gluons. The number of gluons for each case is chosen so that
the cell size in ZPC is no smaller than the interaction length to
ensure the numerical accuracy.

\begin{figure*}[htb]
\centerline{\includegraphics[scale=0.35]{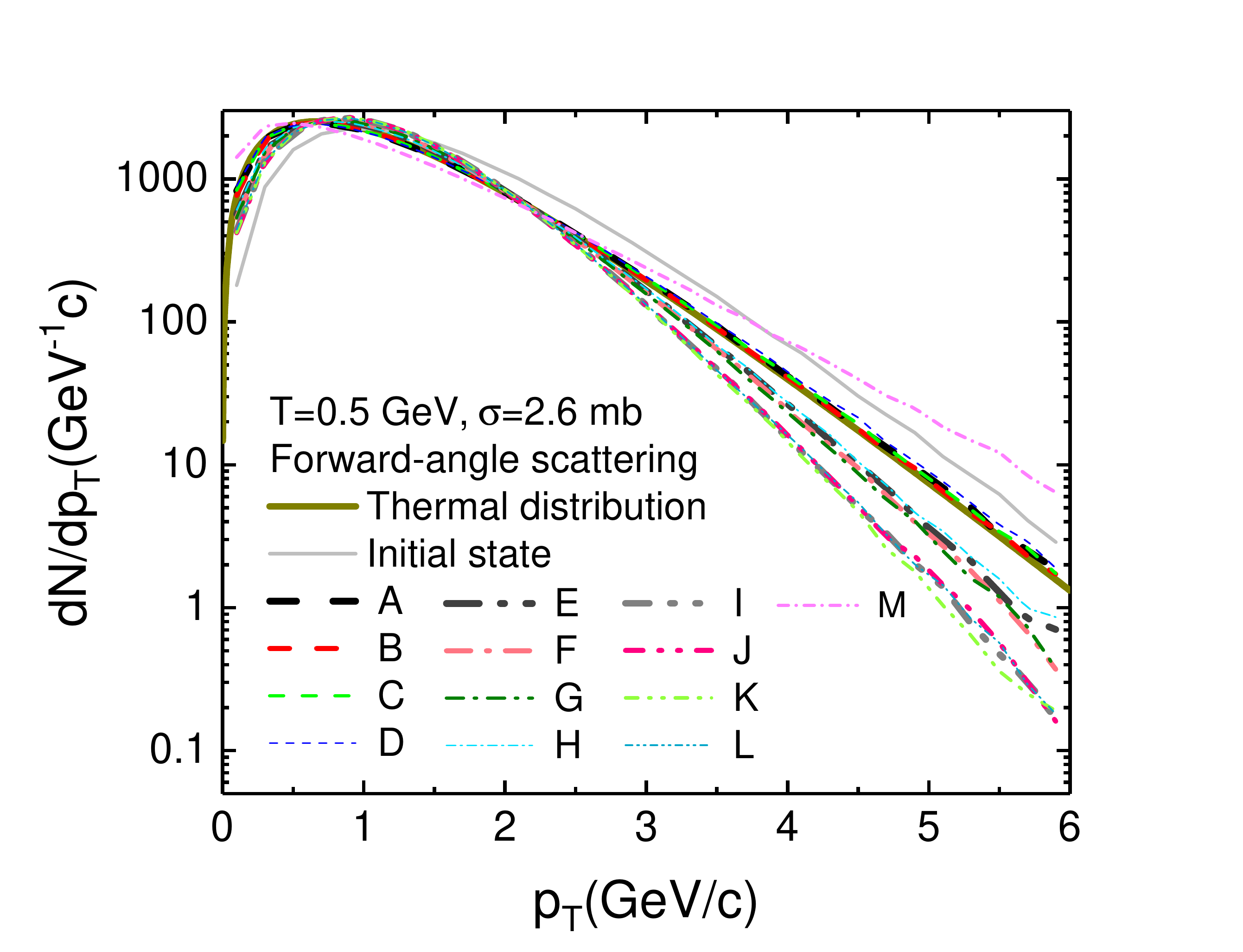}}
\caption{The final $\pt$ distributions from different
  collision schemes for elastic gluon scatterings in a box with
  $T=0.5$ GeV and forward-angle scatterings at $\sigma=2.6$ mb.}
\label{fig:sche}
\end{figure*}

Figure~\ref{fig:sche} shows the final $\pt$ distributions from
different collision schemes for the above case in comparison with the
initial distribution (thin solid curve) and the expected final state
distribution (thick solid curve). Note that each final distribution in
Fig.~\ref{fig:sche} is obtained by running the ZPC parton cascade for
a global time of 6.1 fm$/c$, when the $\pt$ distribution has become
very stable (as can be seen from Fig.\ref{fig:pt2}). We see that the
numerical solution of ZPC depends significantly on the collision
scheme in this case. In addition, the distributions from collision
schemes with the same ordering time (i.e., schemes A to D, or schemes
E to H, or schemes I to L) are relative close to each other.
Collision scheme G, which uses $(ct_1+ct_2)/2$ as both the collision
time and ordering time, is the default collision scheme of
ZPC~\cite{Zhang:1997ej} and also used in the AMPT model
~\cite{Lin:2004en}, so we label it as default ZPC in this study.
We see that the final $\pt$ distribution from the default ZPC scheme in
Fig.~\ref{fig:sche} deviates considerably from the expected thermal
distribution, and so do the results from most collision schemes.
However, the final $\pt$ distribution from collision scheme
B, which uses $min(ct_1,ct_2)$ as both the collision time and ordering
time,  is the closest to the expected
thermal distribution. Therefore we focus on collision scheme B and
call it the new scheme. 

Currently our finding that the collision scheme using time
$min(ct_1,ct_2)$ best preserves the equilibrium momentum 
distribution is a numerical observation. More generally the
ordering time or the collision time in the global frame can be chosen
as a function of $ct_1$ and $ct_2$.  
Indeed we could fine-tune the new collision scheme by choosing a point 
(near $min(ct_1,ct_2)$) on the linear interpolation line between
$ct_1$ and $ct_2$ to preserve the equilibrium momentum distribution
even better. Causality violation usually suppresses the collision
rates, which is the case for the default ZPC scheme as shall be shown
in Fig.~\ref{fig:WT0}. Therefore we can expect that choosing time
$min(ct_1,ct_2)$ instead of $(ct_1+ct_2)/2$ enhances the collision
rates and alleviates the effect of causality violation. 
Other than this, we find no clear theoretical arguments
on why the new scheme works better to suppress the causality
violation. It may be related to correlated functions in
theories such as the Kadanoff-Baym equations~\cite{Lipavsky:1986zz,Semkat}. 
However, the various collision schemes in Table~\ref{schemes} are only
different at finite opacities, where causality violation 
complicates the theoretical analysis of different schemes.

\section{Parton subdivision}
\label{subd}

Naively a parton cascade is only correct in the dilute limit to
preserve causality and Lorentz covariance
~\cite{Zhang:1998tj,Molnar:2000jh,Molnar:2001ux,Cheng:2001dz}, where
the particle range of interaction is much smaller than the mean free
path. Their ratio can be written as \cite{Zhang:1998tj}
\begin{equation}
\chi =\sqrt{\frac{\sigma}{\pi } }/\lambda
=n \sqrt{\frac{\sigma^3}{\pi } },
\label{chi1}
\end{equation}
where $n$ is the parton density and $\lambda$ is the mean free path.
We can use $\chi$ to represent the opacity of the parton system,
and the dilute limit means $\chi \ll 1$.
Above the dilute limit, a parton cascade may suffer from the causality
violation~\cite{Kodama:1983yk,Kortemeyer:1995di,Zhang:1996gb,Zhang:1998tj,Cheng:2001dz},
which is an artifact of the geometrical interpretation of cross
section in the cascade method.
This is why we see the differences in the numerical solutions from difference collision schemes in Fig.~\ref{fig:sche}, which case corresponds to $\chi=2.0$.

The parton subdivision technique~\cite{Pang:1997,Zhang:1998tj}
can be used to reduce the numerical artifact from the causality
violation, and the numerical solution of a parton cascade will be
correct in the limit of large parton subdivision factor.
This is because the Boltzmann equation in Eq.(\ref{be1}) may be expressed as
\begin{equation}
p^\mu \partial_\mu f(\bm r,\bm p,t)\propto \sigma f_1(\bm r_{1},\bm p_{1},t)f_2(\bm r_{2},\bm p_{2},t).
\label{be2}
\end{equation}
Therefore the following transformation keeps the above equation invariant:
\begin{equation}
f(\bm r,\bm p,t) \rightarrow  l \times f(\bm r,\bm p,t),
~~~\sigma  \rightarrow \sigma / l;
\label{subd1}
\end{equation}
but it reduces the opacity as
\begin{equation}
\chi \rightarrow \chi /\sqrt {l}.
\label{chi2}
\end{equation}
In the above, $l$ is the subdivision factor, typically an integer much
greater than one.  Therefore at large enough $l$ the transformed
parton system will reach the dilute limit and thus the cascade
solution for its evolution will be accurate.

We emphasize that the angular distribution of the cross section must
not be changed when performing the above subdivision transformation
Eq.~(\ref{subd1}) to ensure the invariance of the Boltzmann equation;
this can be clearly seen from the term $|\mathcal M|^2$ in Eq.(\ref{be1}).
Therefore the exact
transformation for parton
subdivision is the following:
\begin{equation}
f(\bm r,\bm p,t) \rightarrow  l \times f(\bm r,\bm p,t),
~~~\frac{d\sigma}{d \hat t}  \rightarrow \frac{d\sigma}{d \hat t} / l.
\label{subd2}
\end{equation}
This is especially relevant for forward-angle scatterings, because
there the total cross section as well as the angular distribution
are determined by the screening mass $\mu$ as shown in
Eqs.(\ref{dsigmadt}-\ref{sigma}). When parton subdivision requires the
decrease of the forward-angle cross section of Eq.(\ref{sigma}), one
should not do that by increasing $\mu$ by a factor of $\sqrt{l}$ because
that would change the angular distribution of the scatterings. Instead
one can decrease the $\alpha_s$ parameter by a factor of $\sqrt{l}$ in
Eqs.(\ref{dsigmadt}-\ref{sigma}), which decreases the total scattering
cross section while keeping its angular distribution.

In the standard subdivision method one increases the initial parton
number per event by factor $l$ while decreasing the cross section by
the same factor. This method can be schematically represented by the
following transformation:
\begin{equation}
N \rightarrow l \times N, V {~\rm unchanged},
\label{subd3}
\end{equation}
where $N$ is the initial parton number in an event and $V$ is the
initial volume of the parton system.
For box calculations of elastic scatterings in this study,
of course the parton number in an event and the volume
do not change with time.
Since the number of possible collisions scales with $l^2$,
the subdivision method is very expensive in terms of the computation
time, which roughly scales with $l^2$ per subdivision event or with
$l$ per simulated parton.
However, for box calculations where the density function $f(\bm r,\bm
p,t)$ is spatially homogeneous, we can realize parton subdivision with
a different method. This new subdivision method can be schematically
represented by
\begin{equation}
N {~\rm unchanged}, V \rightarrow V/l,
\label{subd4}
\end{equation}
where we decrease the volume of the box by factor $l$ while keeping
the same parton number and momentum distribution in each event.
Because the parton number per event does not change, this subdivision
method is much more efficient than the standard subdivision method,
therefore we can afford a huge subdivision factor such as $10^6$
(instead of the usual value of up to a few hundreds).
For all parton subdivision calculations in this study,
we shall use this new subdivision method with $l=10^6$
under the default ZPC scheme (unless specified otherwise).

\begin{figure*}[htb]
  \begin{minipage}[t]{0.333\linewidth}
    \includegraphics[width=1.15\textwidth]{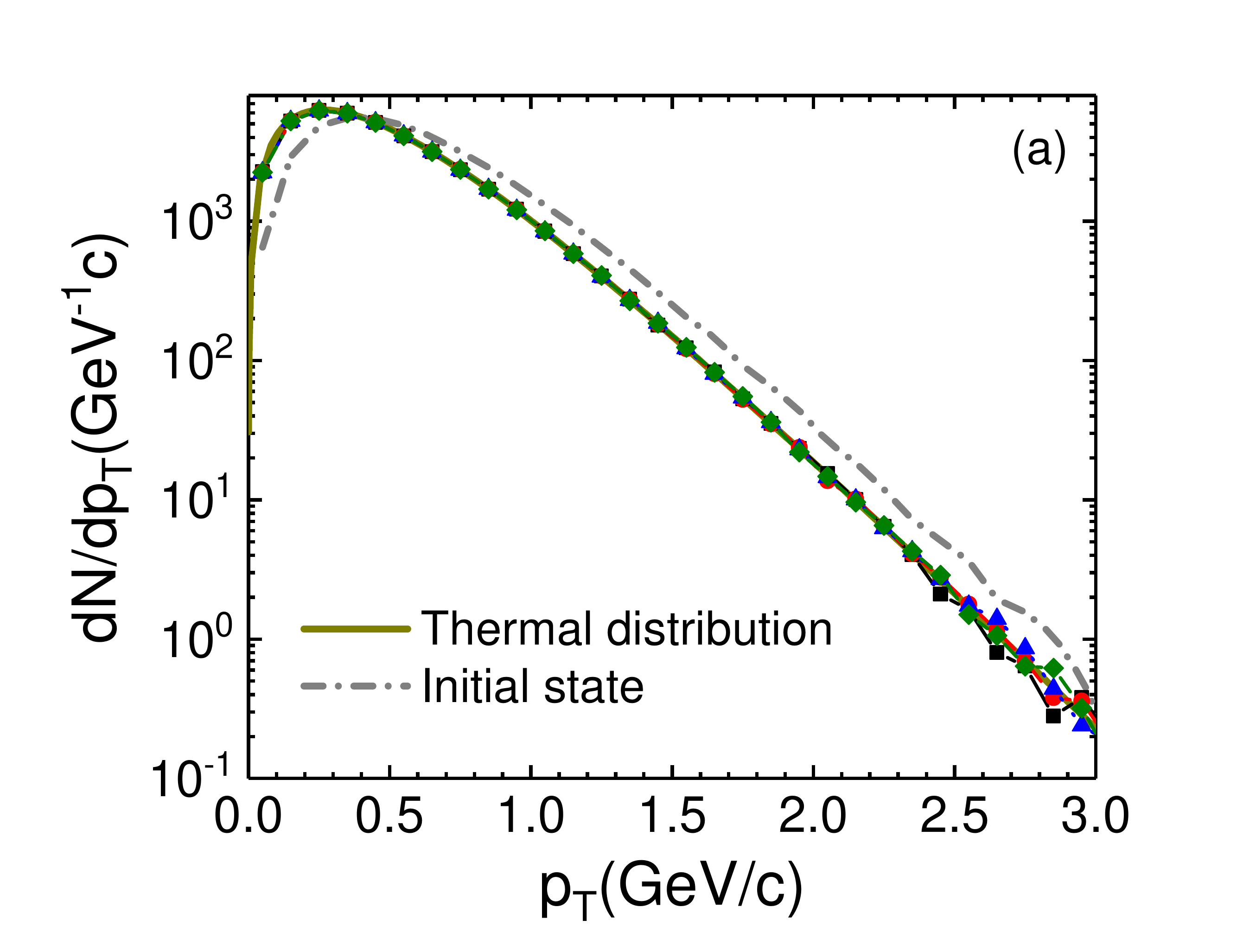}
  \end{minipage}%
  \begin{minipage}[t]{0.333\linewidth}
    \centering
    \includegraphics[width=1.15\textwidth]{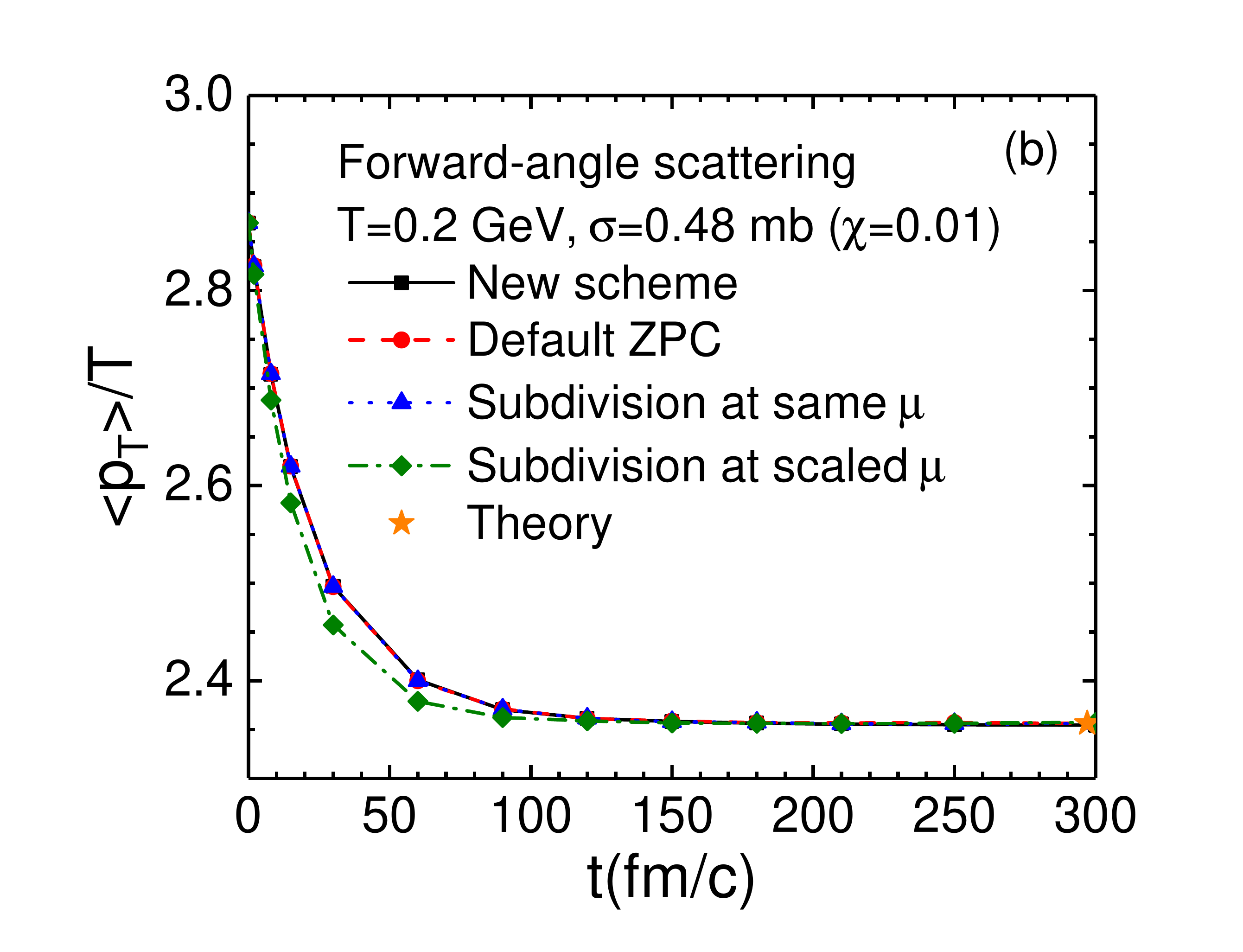}
  \end{minipage}%
  \begin{minipage}[t]{0.333\linewidth}
    \centering
    \includegraphics[width=1.15\textwidth]{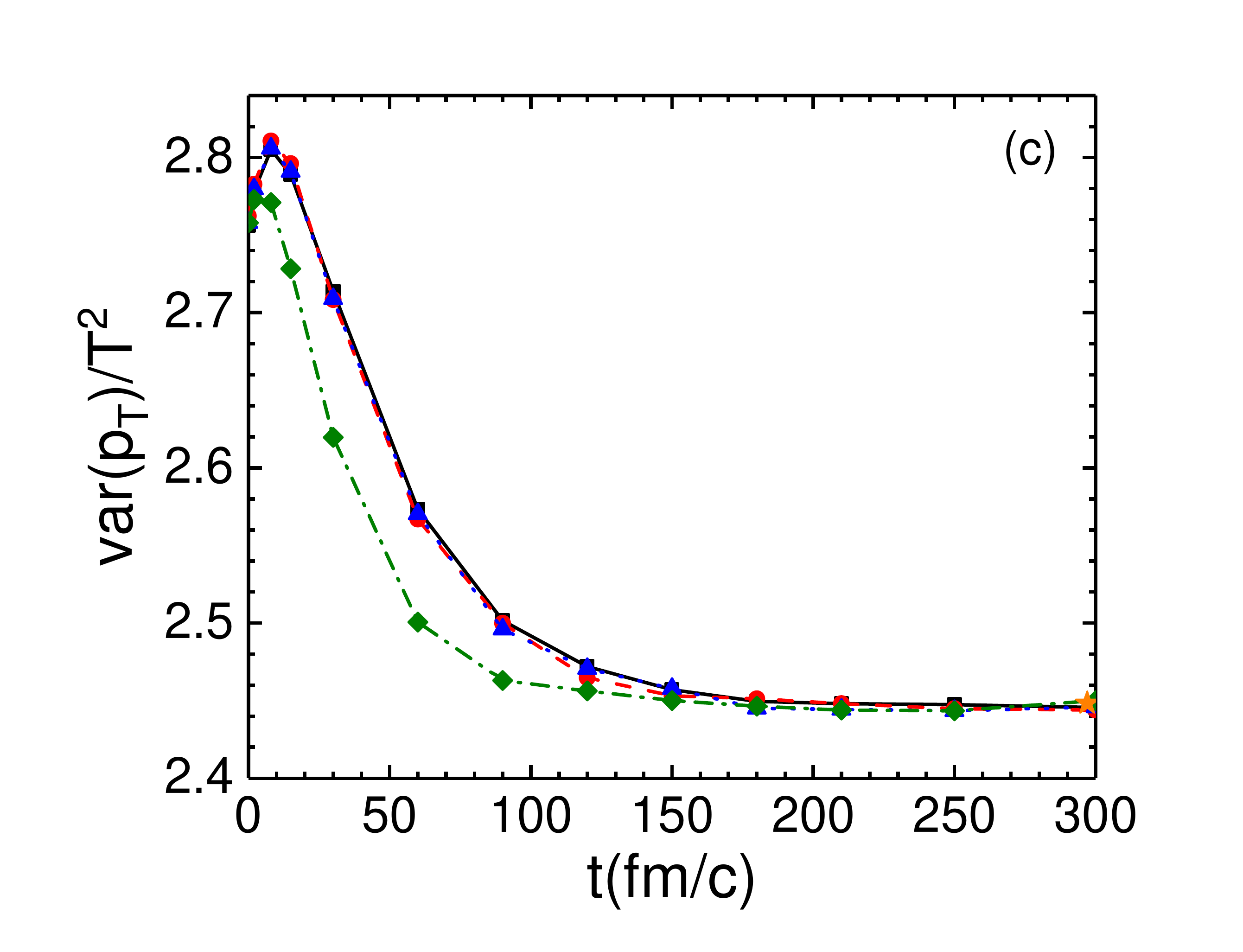}
  \end{minipage}
\caption{(a) The final $\pt$ distribution, (b) time
  evolution of $\langle \pt \rangle/T$, and (c) time evolution of
  var($\pt$)$/T^2$ from the new scheme, the default ZPC scheme, and
  two parton subdivision methods in a low-opacity test at
  $\alpha_{s}=0.20$, $T=0.2$ GeV for forward-angle scatterings at
  $\sigma=0.48$ mb. The star symbols represent the theoretical values
  at late times.}
\label{fig:pt0}
\end{figure*}

To explicitly show the importance of keeping the same scattering
angular distribution when implementing the parton subdivision method, as
shown by Eq.(\ref{subd2}), we apply the ZPC parton cascade to a
dilute limit. For this case we simulate for each event 4,000 gluons at
$T=0.2$ GeV with an off-equilibrium momentum distribution. We set
$\mu=3.47$ fm$^{-1}$ and $\alpha_s=0.201$, which
then gives a forward-angle scattering cross section
$\sigma=0.48$ mb and $\chi=0.01$ (a dilute system).
Figure~\ref{fig:pt0} shows the results of
the final $\pt$ distribution, time evolution of 
$\langle \pt \rangle/T$, 
and time evolution of var($\pt$)$/T^2$ from different
cascade methods, where
$\langle \pt\rangle$ is the mean transverse momentum of each parton
and
\begin{equation}
{\rm var}(\pt) = \langle {\pt^2} \rangle-\langle \pt \rangle^2
\end{equation}
is the variance of the final $\pt$ distribution.
The results include those from the new scheme, the default ZPC
scheme, the parton subdivision method at $l=10^6$ with an
unchanged scattering angular distribution (i.e., by decreasing
$\alpha_s$ while keeping $\mu$ the same),
and the parton subdivision method at $l=10^6$ with a changed
scattering angular distribution  (i.e., by increasing $\mu$).
Note that these two parton subdivision calculations are performed with
the default ZPC scheme; however, the choice of schemes no longer
affects the numerical results here because the large $l$ value for the
parton subdivision has essentially eliminated the causality
violation.

We first see in Fig.~\ref{fig:pt0} that the results from
the new scheme (solid curves) and the default ZPC scheme  (dashed
curves) agree with each other very well; this is because the effect of
causality violation and thus the dependence on the collision scheme
is very small in this dilute limit.
We also see that they agree with the parton subdivision method
that keeps the same scattering angular distribution (dotted curves)
but their time evolutions disagree with the parton subdivision method
that changes the scattering angular distribution (dot-dashed curves),
thus verifying the parton subdivision method of Eq.(\ref{subd2}).
From now on we shall use this method for parton subdivision for
all the remaining calculations.
In addition, the time evolutions of $\langle \pt\rangle$ and
${\rm var}(\pt)$ are both faster for the parton subdivision method
that changes the scattering angular distribution; this is because
the subdivision scaling $\mu \rightarrow \sqrt{l}~\mu$ used in this
``wrong'' subdivision method makes the angular distribution more
isotropic and thus leads to a higher transport cross section
\cite{Molnar:2001ux} than the ``correct''
subdivision method (dotted curves).
The star symbols in Figs.~\ref{fig:pt0}(b) and \ref{fig:pt0}(c) represent the
theoretical values at late times (or in equilibrium) for the mean
value (scaled by $1/T$) and the variance (scaled by $1/T^2$) of the
$\pt$ distributions, respectively, where
\begin{equation}
\langle \pt \rangle = \frac{3\pi T}{4},
~~{\rm var}(\pt)=\left (8-\frac{9\pi^2}{16} \right ) T^2.
\label{theory}
\end{equation}
At late times all four ZPC calculations in Fig.~\ref{fig:pt0}
reach the correct equilibrium values for this dilute case.

\begin{figure*}[htb]
\centerline{\includegraphics[scale=0.35]{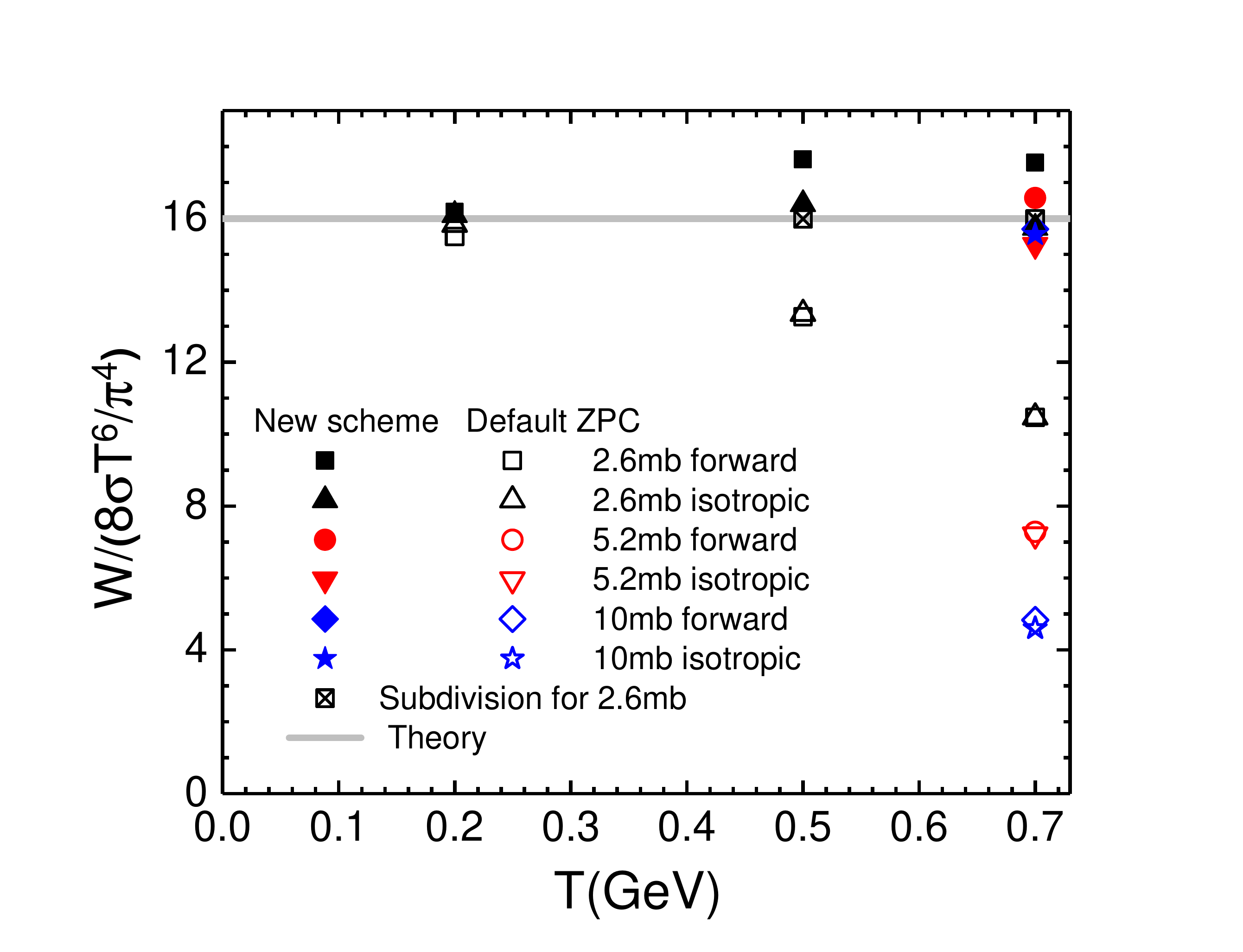}}
\caption{Scaled collision rates per unit volume $W$ as
  functions of temperature $T$ for forward-angle scatterings and
  isotropic scatterings at different cross sections. Results from the
  new scheme and the default ZPC scheme are compared with the parton
  subdivision results and the theoretical value.}
\label{fig:WT0}
\end{figure*}

We also check the collision rates per volume~\cite{Zhang:1998tj}
in Fig.~\ref{fig:WT0}, which shows the results for cases with
different cross sections and temperatures for both
forward-angle and isotropic scatterings.
The horizontal line represents the (scaled) expected rate per volume
for a massless gluon system in equilibrium, which is given by~\cite{Zhang:1998tj}
\begin{equation}
W=\frac{8\sigma T^6}{\pi ^4}F \left (\frac{2m}{T} \right )
 \textrm{~~~with~~~} F(x)=\int_{x}^{\infty }dyy^2(y^2-x^2)K_{1}(y),
\label{w2}
\end{equation}
In the above, $K_{1}(y)$ is the modified Bessel function, and
$F(0)=16$ for massless gluons that we consider in this study.
We see that as expected the small-opacity results (symbols at $T=0.2$ GeV) from
the new scheme and the default ZPC scheme for both forward-angle and
isotropic scatterings are almost the same.
For large opacities, however, the results (symbols at $T=0.5$ GeV and
$0.7$ GeV) depend significantly on the collision scheme; they also
depend on the scattering angular distribution in some cases.
In particular,  the collision rate per volume from the default ZPC
scheme gets much  lower than the theoretical expectation for large
cross sections or parton densities (which scales as $T^3$).
On the other hand, the collision rates per volume from the new scheme
are much closer to the theoretical value, even at large opacities.
Also, it is no surprise that the parton subdivision results agree
with the theoretical expectation. 
Note that the collision rates per unit volume are essentially the same
when we use the equilibrium initial condition instead of the
off-equilibrium initial condition.

\section{Main results and discussions}
\label{results}

\subsection{The $\pt$ distribution and its time evolution}
\label{resultsA}

We now use different cases to test the new collision scheme in
comparison with the default ZPC scheme and exact results on the
momentum distribution.
As noted before, the initial momentum distribution in each of these
calculations is off-equilibrium so that we can better observe the
time evolution and equilibration of the momentum distribution.

\begin{figure*}[htb]
  \begin{minipage}[t]{0.333\linewidth}
    \includegraphics[width=1.15\textwidth]{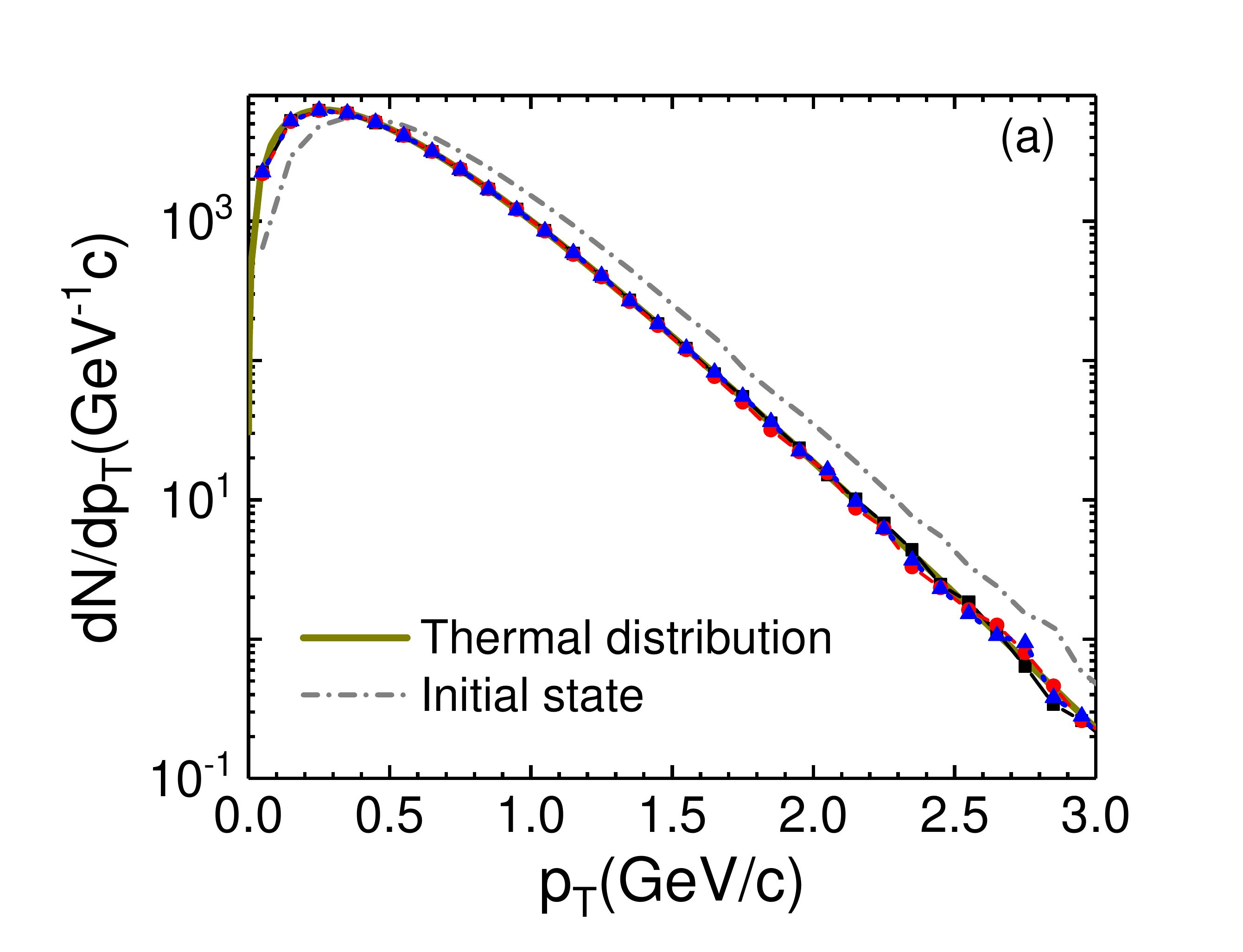}
  \end{minipage}%
  \begin{minipage}[t]{0.333\linewidth}
    \centering
    \includegraphics[width=1.15\textwidth]{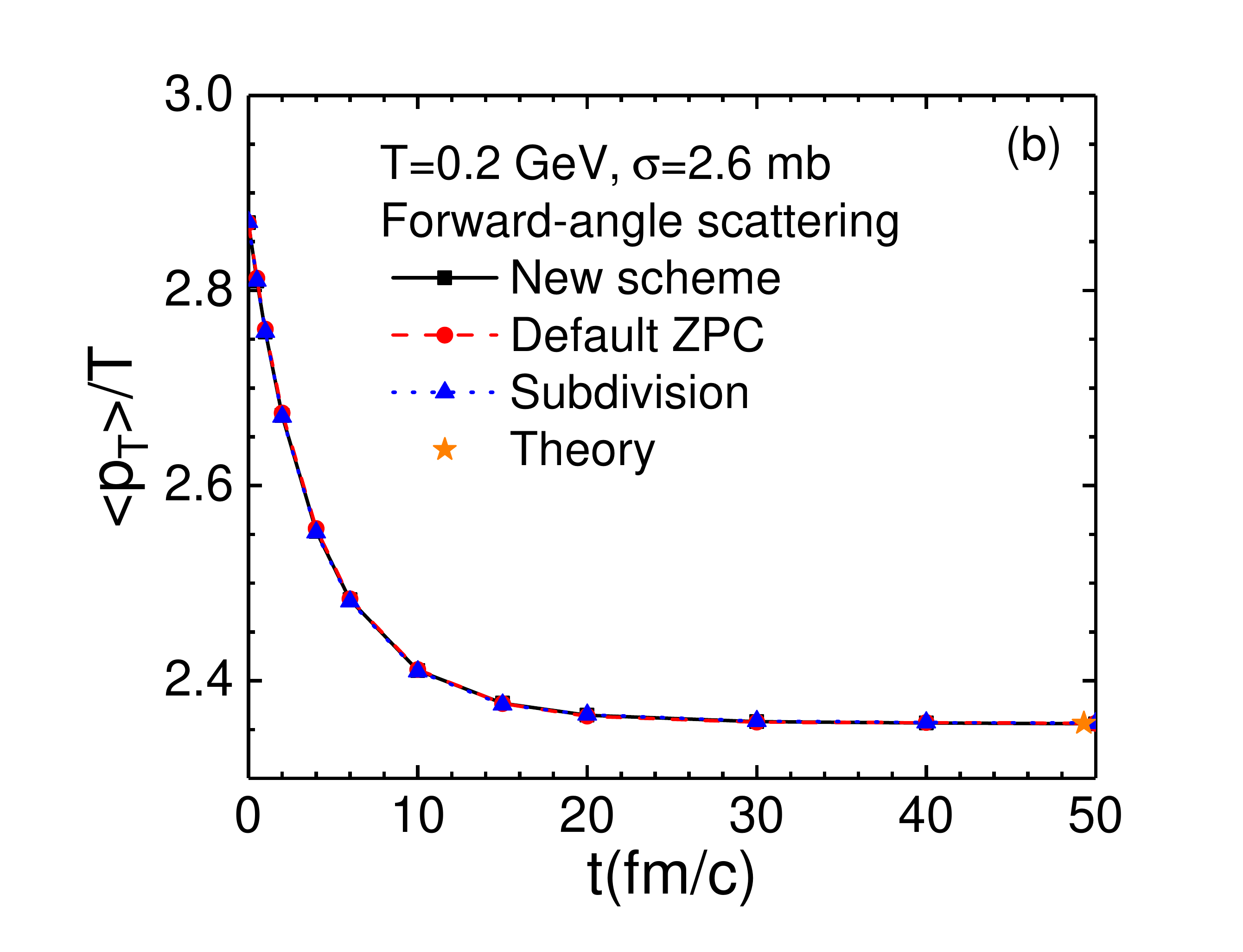}
  \end{minipage}%
  \begin{minipage}[t]{0.333\linewidth}
    \centering
    \includegraphics[width=1.15\textwidth]{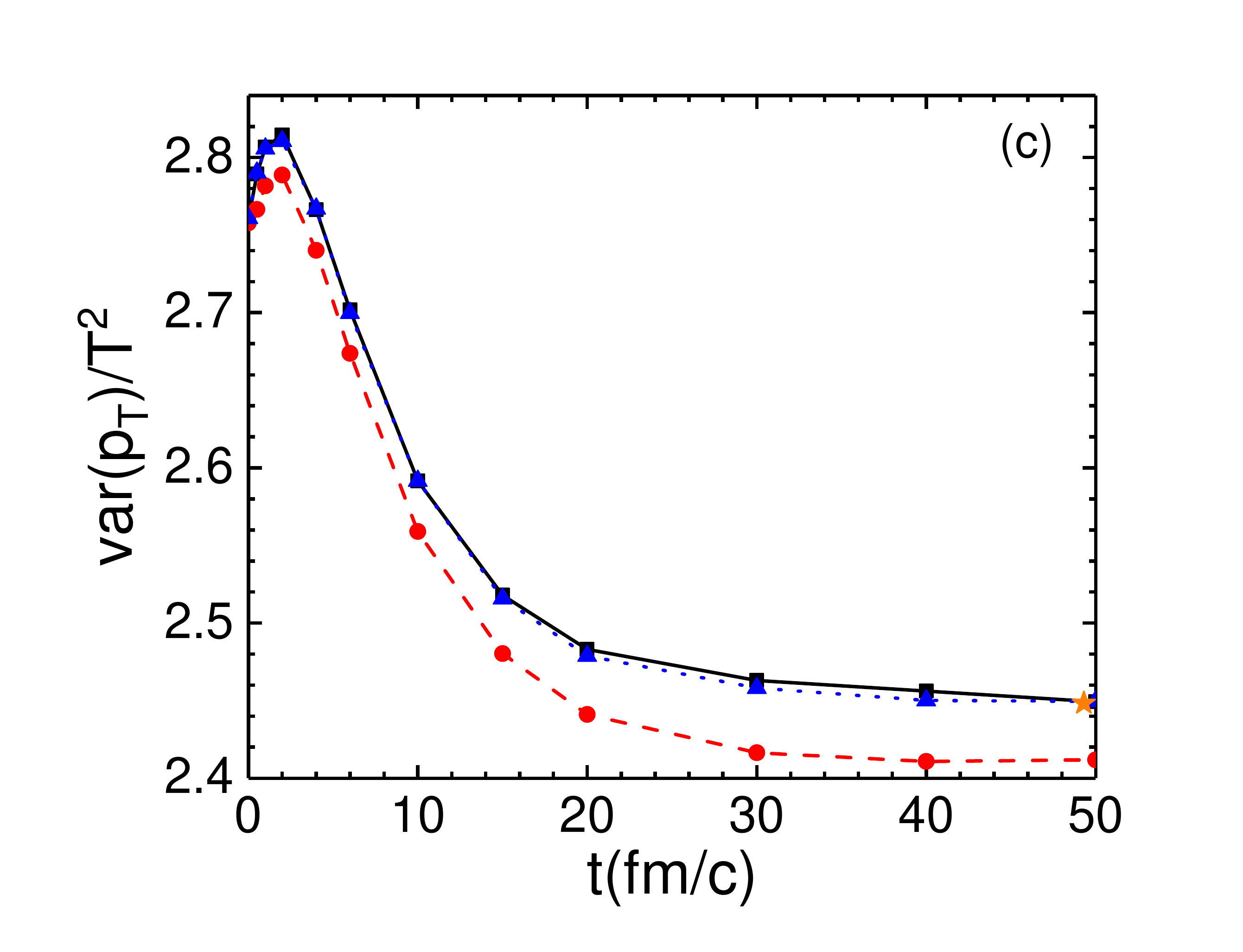}
  \end{minipage}
\caption{(a) The final $\pt$ distributions, (b) time
  evolutions of $\langle \pt \rangle/T$, and (c) time evolutions
  of var($\pt$)$/T^2$ from the new scheme and the default ZPC scheme
in comparison with the parton subdivision results in a box at $T=0.2$ GeV for forward-angle scatterings at $\sigma=2.6$ mb ($\chi=0.13$).}
\label{fig:pt1}
\end{figure*}

We start with a test of forward-angle scatterings at low opacity,
where $\sigma=2.6$ mb and $T =0.2$ GeV that correspond to
$\chi=0.13$.
Figure~\ref{fig:pt1} shows the final $\pt$ distributions,
the time evolutions of $\langle \pt\rangle/T$, and the time evolutions
of var($\pt$)$/T^2$.
In Fig.~\ref{fig:pt1}(a) we see  no obvious difference between the
final $\pt$  distributions of the three methods of doing the parton
cascade, i.e., the new scheme, the default ZPC scheme, and the
parton subdivision method at $l=10^6$.
They are also consistent with the thermal distribution, which is
expected because the small $\chi$ value here means that
the effect from causality violation should be quite small.
Furthermore, the time evolutions of the mean transverse momentum in
Fig.~\ref{fig:pt1}(b) also show little difference among the three
methods.
However, we observe some difference in the time evolutions of
the variance of the $\pt$  distributions in Fig.~\ref{fig:pt1}(c);
in particular, the variance from the default ZPC scheme is
obviously smaller than the other two results soon after the start of
parton cascade, meaning that the $\pt$ distribution from the default
ZPC scheme is somewhat narrower in width, even at late times.
In addition, Figs.~\ref{fig:pt1}(b) and \ref{fig:pt1}(c) indicate that
the $\pt$ distributions from the new scheme and from the parton
subdivision method follow a similar time evolution and at late times
they agree with the theoretical expectations (star symbols).

\begin{figure*}[htb]
  \begin{minipage}[t]{0.333\linewidth}
    \includegraphics[width=1.15\textwidth]{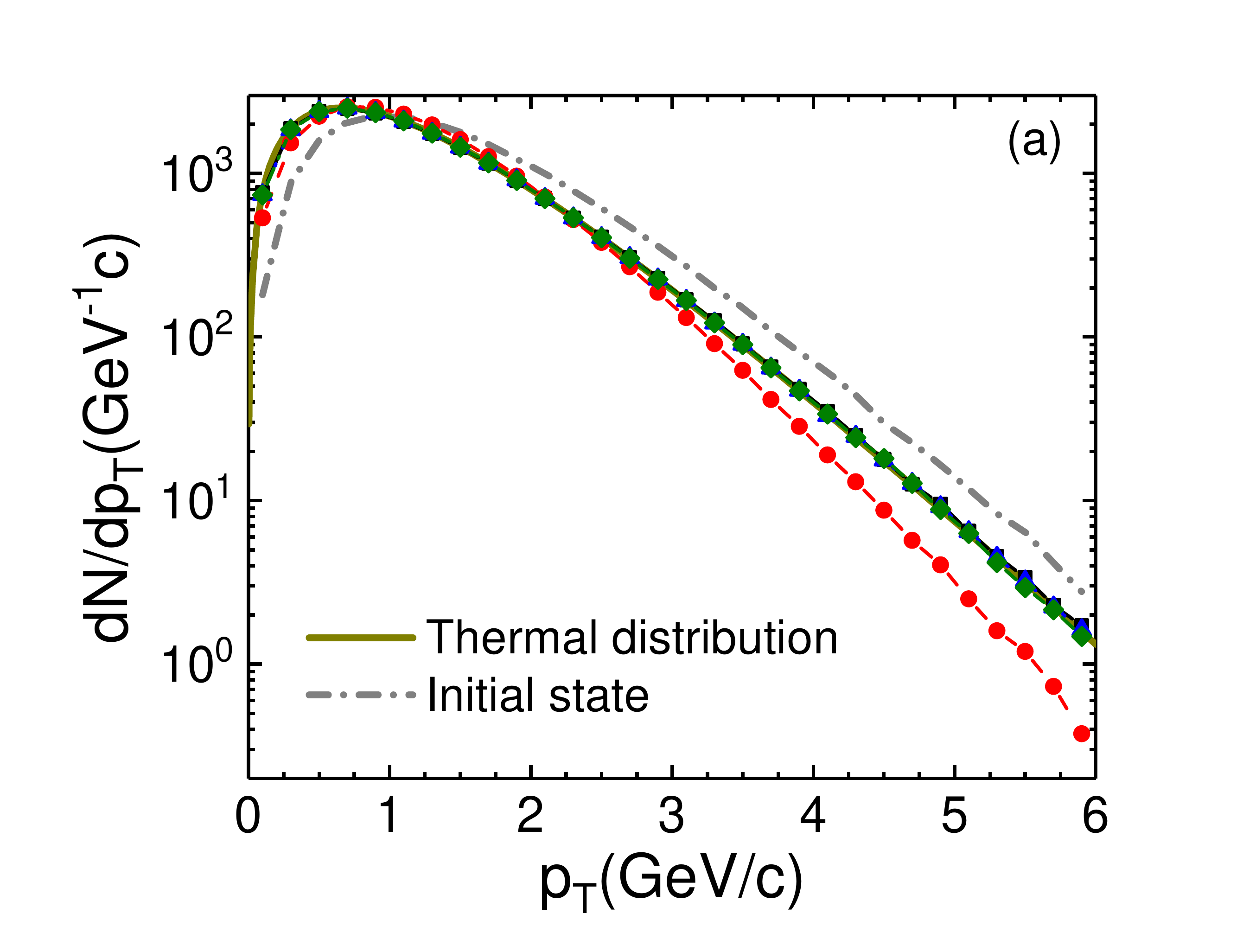}
  \end{minipage}%
  \begin{minipage}[t]{0.333\linewidth}
    \centering
    \includegraphics[width=1.15\textwidth]{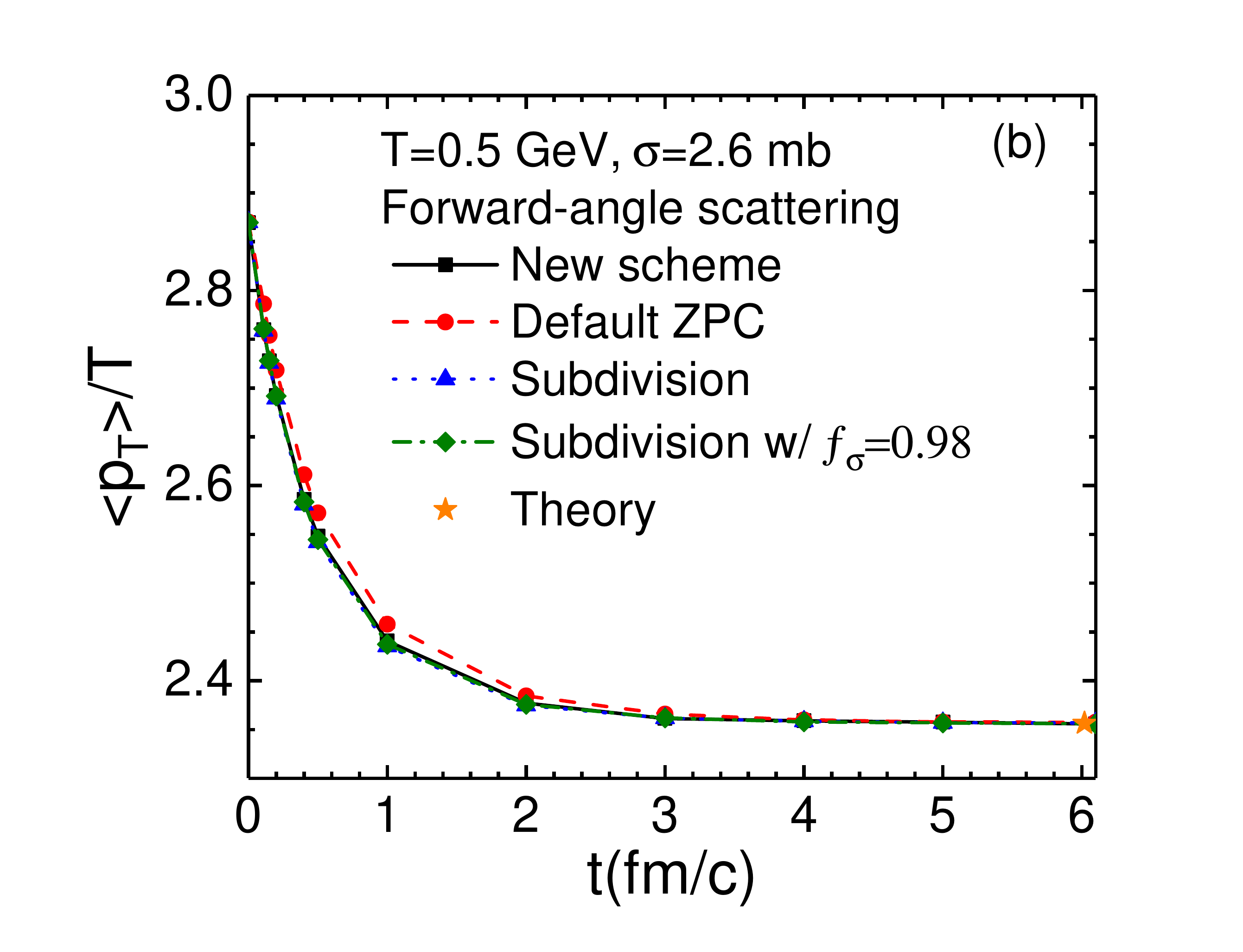}
  \end{minipage}%
  \begin{minipage}[t]{0.333\linewidth}
    \centering
    \includegraphics[width=1.15\textwidth]{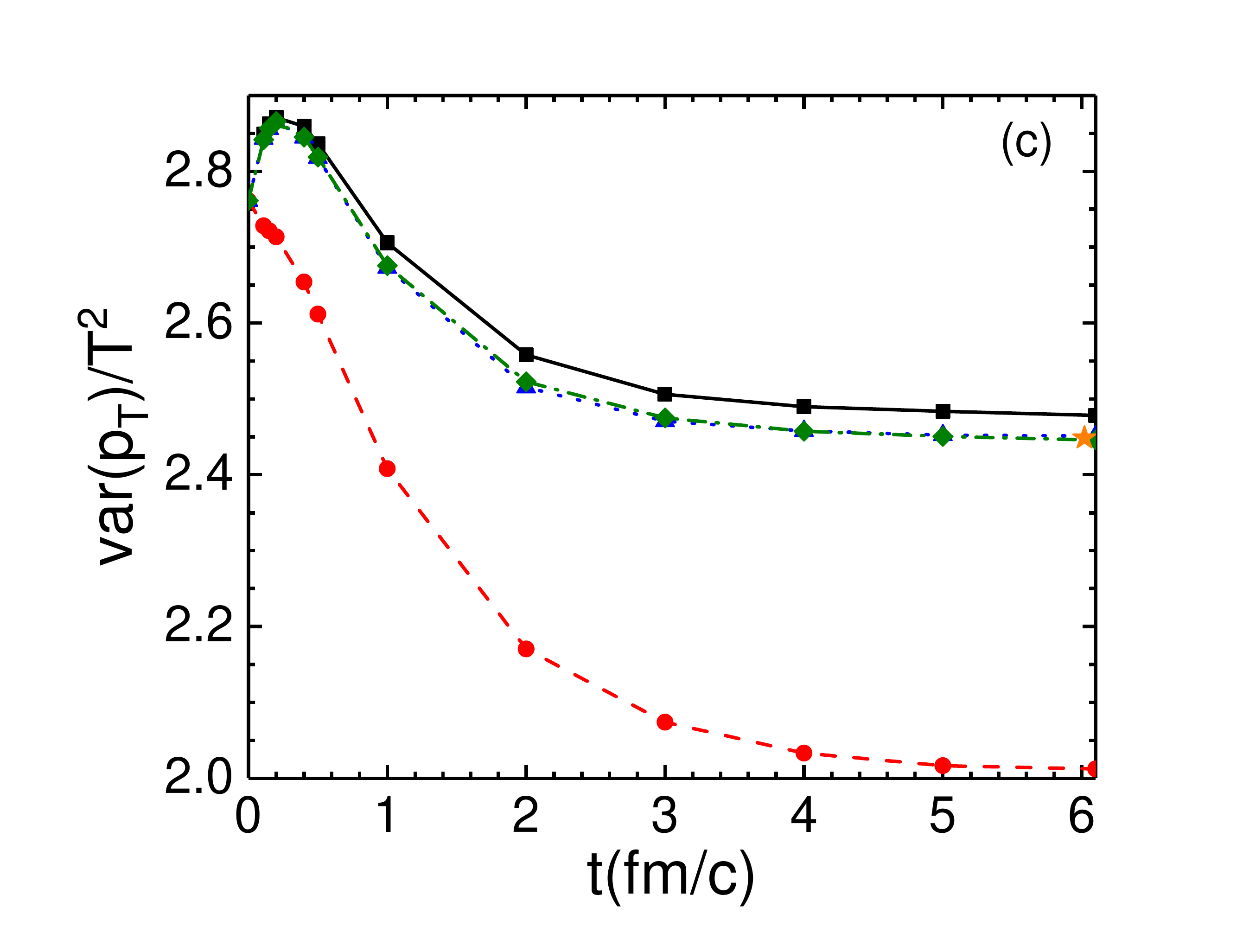}
  \end{minipage}
\caption{Same as Fig.~\ref{fig:pt1} but at $T=0.5$ GeV for
  forward-angle scatterings at $\sigma=2.6$ mb ($\chi=2.0$).}
\label{fig:pt2}
\end{figure*}

\begin{figure*}[htb]
  \begin{minipage}[t]{0.333\linewidth}
    \includegraphics[width=1.15\textwidth]{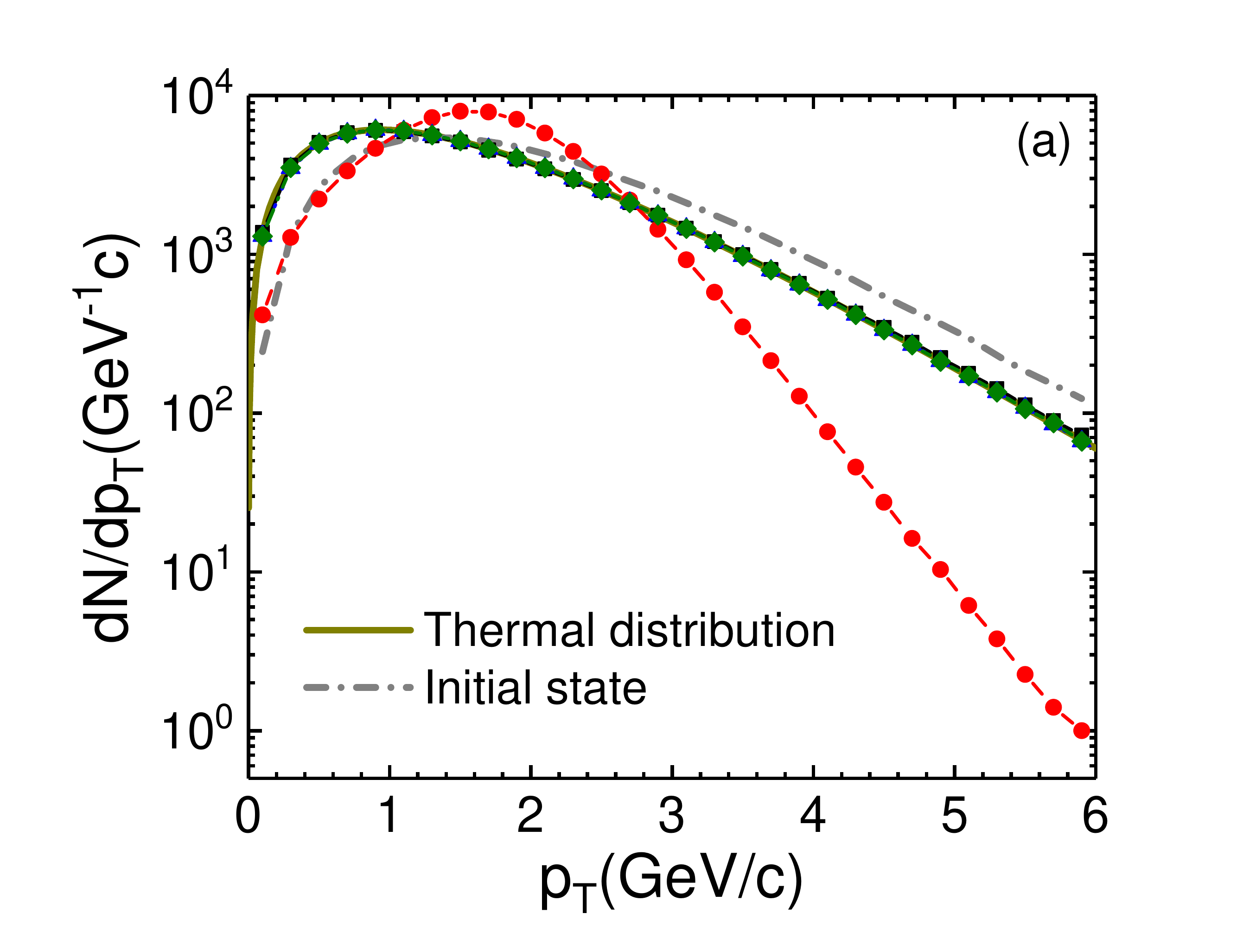}
  \end{minipage}%
  \begin{minipage}[t]{0.333\linewidth}
    \centering
    \includegraphics[width=1.15\textwidth]{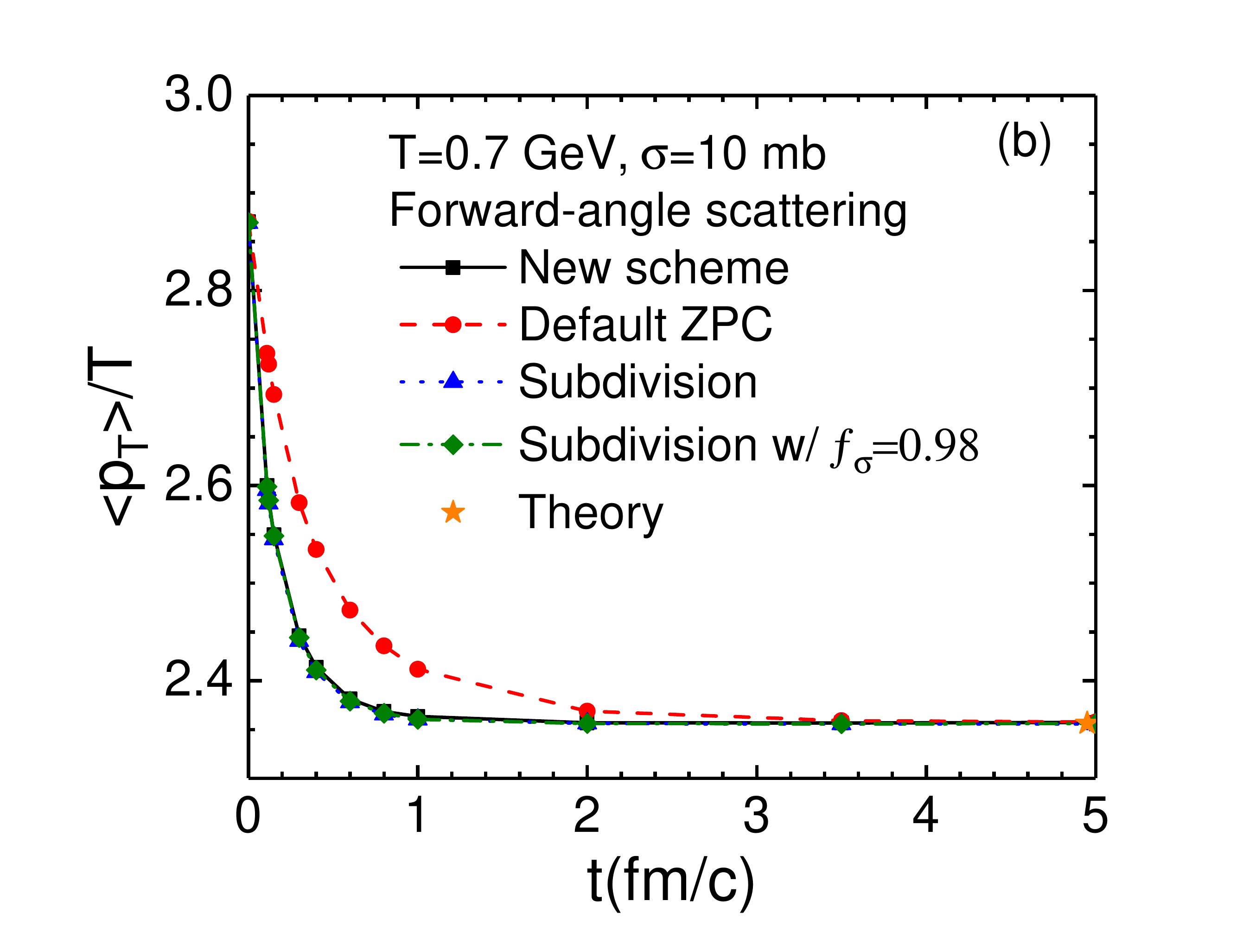}
  \end{minipage}%
  \begin{minipage}[t]{0.333\linewidth}
    \centering
    \includegraphics[width=1.15\textwidth]{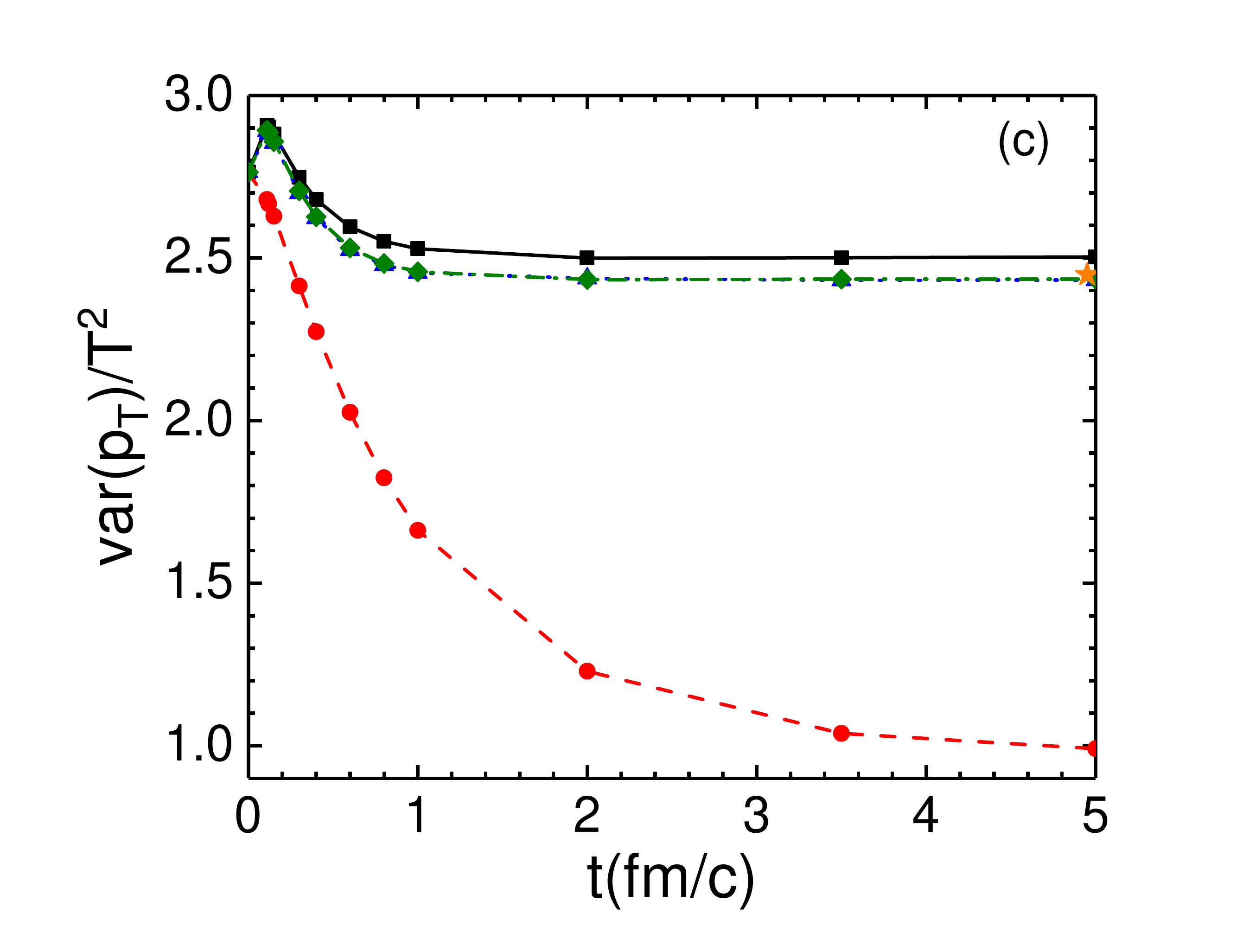}
  \end{minipage}
\caption{Same as Fig.~\ref{fig:pt1} but at $T=0.7$ GeV for
  forward-angle scatterings at $\sigma=10$ mb ($\chi=41$).} 
\label{fig:pt3}
\end{figure*}

The results of two cases of higher opacities, one at $T=0.5$ GeV and
$\sigma=2.6$ mb, another  at $T=0.7$ GeV and $\sigma=10$ mb,
for forward-angle scatterings are shown in
Fig.~\ref{fig:pt2} and Fig.~\ref{fig:pt3}, respectively.
The first case as shown in Fig.~\ref{fig:pt2} corresponds to $\chi=2.0$;
we see that the $\pt$ distribution and its variance
from the default ZPC scheme both deviate significantly from the
``exact'' parton subdivision results, although the time evolutions of
$\langle \pt\rangle$ are close to each other.
On the other hand, results from the new scheme are very close to the
parton subdivision results, which agree with theoretical
expectations at late times.
The second case as shown in Fig.~\ref{fig:pt3} corresponds to
$\chi=41$, which serves as an example of extreme opacity.
We see qualitatively the same features as seen in Fig.~\ref{fig:pt2},
but the results from the default ZPC scheme
are now much further away from the parton subdivision results, including
its time evolution of $\langle \pt\rangle$. Again, results from the
new scheme are quite close to the subdivision results or the
theoretical expectations even at this extreme opacity.

\begin{figure*}[htb]
  \begin{minipage}[t]{0.333\linewidth}
    \includegraphics[width=1.15\textwidth]{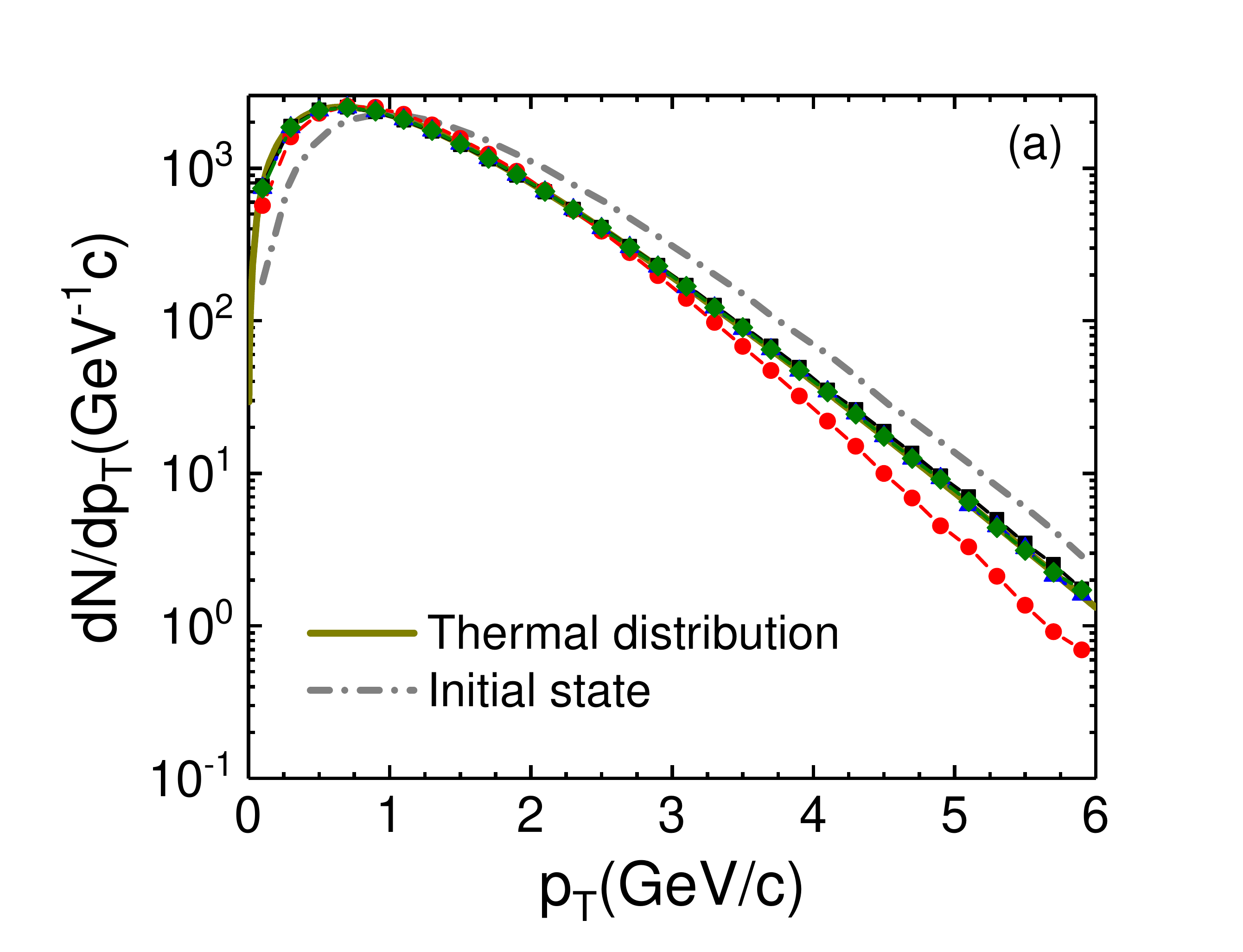}
  \end{minipage}%
  \begin{minipage}[t]{0.333\linewidth}
    \centering
    \includegraphics[width=1.15\textwidth]{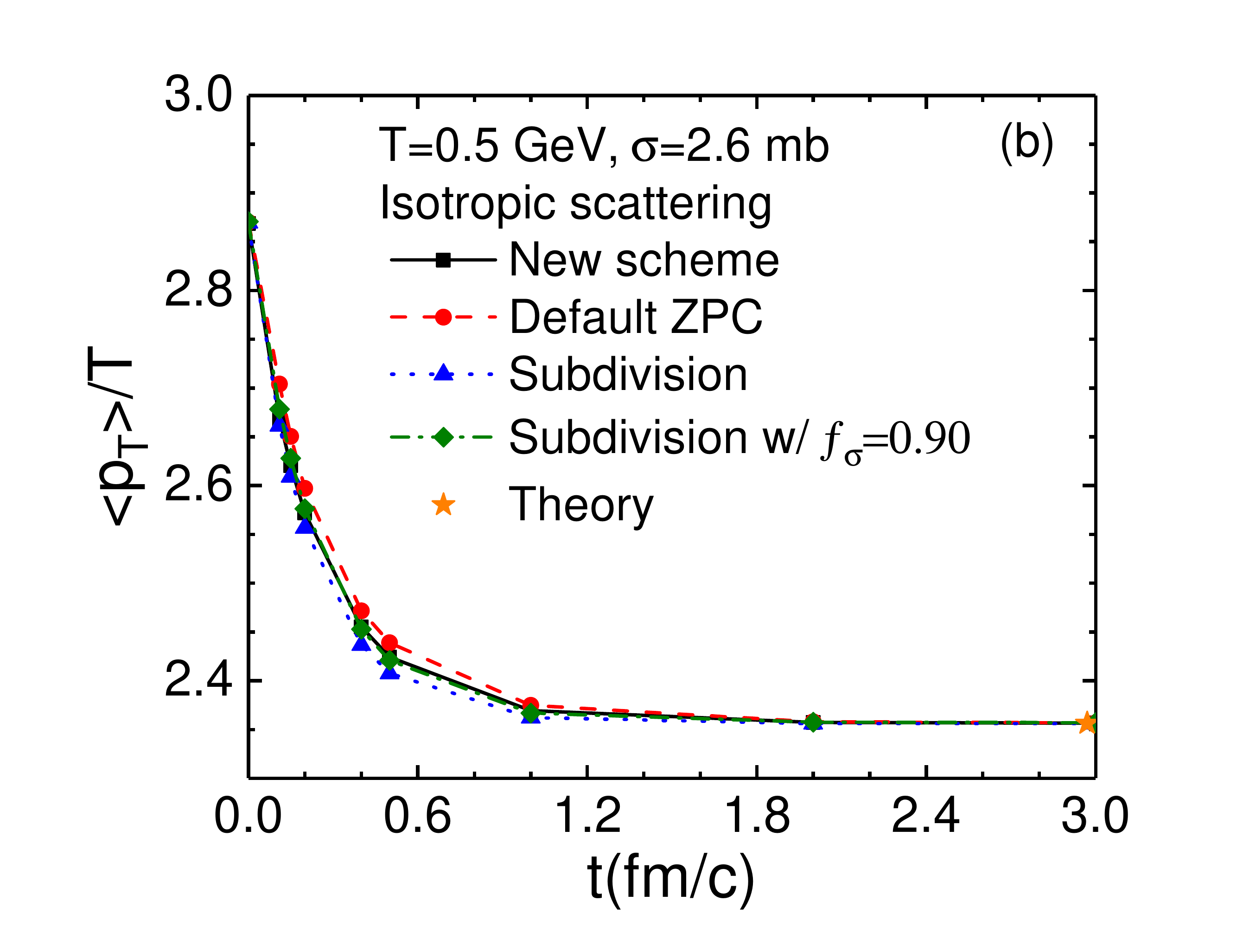}
  \end{minipage}%
  \begin{minipage}[t]{0.333\linewidth}
    \centering
    \includegraphics[width=1.15\textwidth]{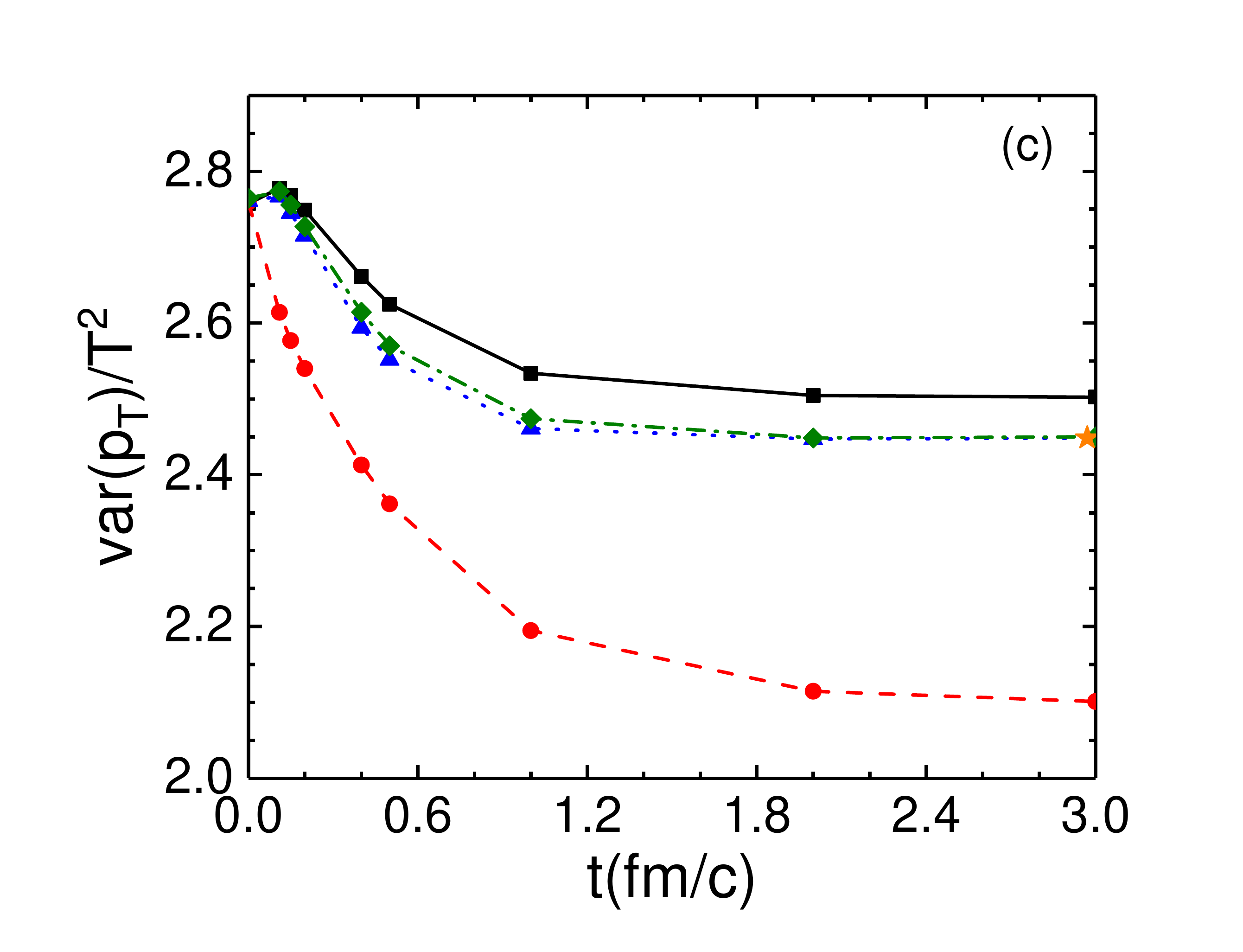}
  \end{minipage}
\caption{Same as Fig.~\ref{fig:pt2} but for isotropic scatterings.}
\label{fig:pt4}
\end{figure*}

We have also tested isotropic scatterings and reached similar
conclusions. As an example, Fig.~\ref{fig:pt4} shows the results for
isotropic scatterings for the case of $T=0.5$ GeV and
$\sigma=2.6$ mb (i.e., $\chi=2.0$).
We see the same features as those shown in Fig.~\ref{fig:pt2}
for forward-angle scatterings, e.g.,
the results of the new scheme are close to the subdivision results
while the default ZPC scheme gives very different results that are
far from the theoretical expectations at late times.
Therefore we conclude that for box calculations the new collision scheme
(i.e., scheme B in Table~\ref{schemes}) is very accurate over a large
range of opacities and much better than the default ZPC collision
scheme.

\begin{figure*}[htb]
\centerline{\includegraphics[scale=0.35]{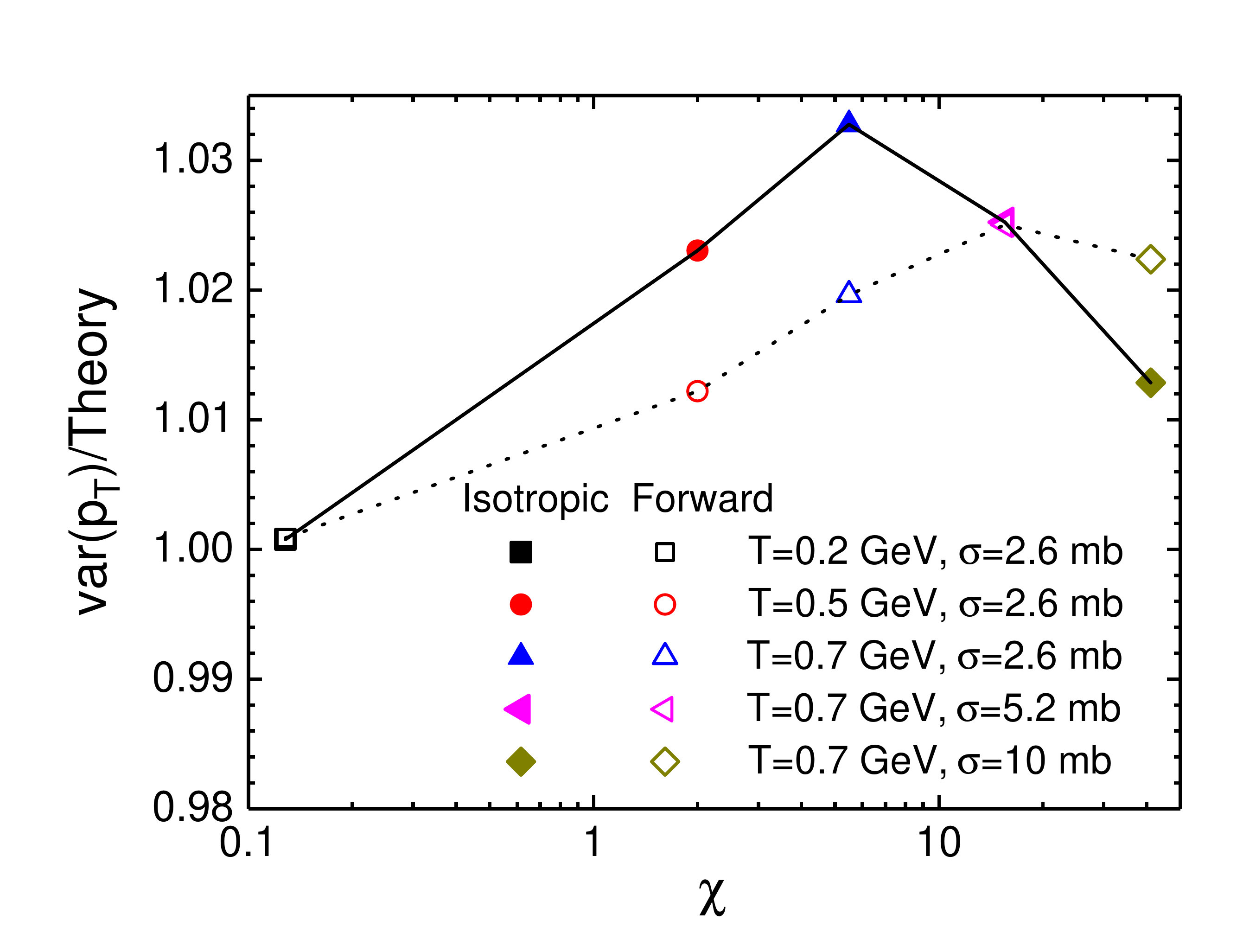}}
\caption{The ratio between the final var($\pt$) from the new
  scheme and the theoretical value versus the opacity parameter $\chi$
  for different cases.}
\label{fig:chi}
\end{figure*}

To characterize the accuracy of the final $\pt$ distribution from the
new collision scheme, we may compare the final $\langle \pt\rangle/T$
and ${\rm var}(\pt)/T^2$ with the corresponding theoretical values in
Eq.~(\ref{theory}).
However, we can see from the figures that the final $\langle \pt
\rangle$ value at late times  from every box calculation in this study
agrees with  Eq.~(\ref{theory}); this is due to the momentum isotropy
in equilibrium and the energy conservation because the average energy per
parton ($3T$) does not depend on the collision scheme or method.
Therefore we choose to use the ratio between the final
${\rm var}(\pt)/T^2$ value and the theoretical value to represent the
accuracy of the new collision scheme. The values of this ratio for
different cases  are shown in Fig.~\ref{fig:chi} as functions of the
opacity parameter $\chi$.
We see that the ratio is essentially unity at low opacities, indicating
that there the new scheme is very accurate as expected.
At moderate to high opacities, the deviations of the variance of the
$\pt$ distribution are quite small, up to about 3\%.
Also, an interesting feature for isotropic scatterings is that
the maximum deviation in the variance does not occur at the highest
opacity shown but at a moderate opacity.

We see from Figs.~\ref{fig:pt2}-\ref{fig:pt4} that the time
evolutions of $\langle \pt \rangle$ from the new scheme (dashed curves)
are somewhat different from the parton subdivision results,
although the $\langle \pt \rangle$ values at late times agree well
with the theoretical value of Eq.~(\ref{theory}).
Therefore we may ask the question:
at what equivalent cross section will the parton subdivision method
give the same time evolution of $\langle \pt \rangle$ as the new scheme?
For example, Fig.~\ref{fig:pt4}(b) shows that the time evolution from
the new scheme (solid curve) is slower than the subdivision result at
the same cross section of 2.6 mb (dotted curve); thus we expect
that the subdivision result at a smaller cross section could better
match the time evolution of the new scheme. For this purpose, we can
write schematically a new subdivision transformation:
\begin{equation}
f(\bm r,\bm p,t) \rightarrow  l \times f(\bm r,\bm p,t),
~~~\frac{d\sigma}{d \hat t}  \rightarrow f_\sigma \times
\frac{d\sigma}{d \hat t} / l,
\label{subd5}
\end{equation}
where $f_\sigma$ is the effectiveness factor of cross section.
We then determine the $f_\sigma$ factor by minimizing the difference
between the time evolution from the new scheme and that from the above
parton subdivision method.
More specifically, we minimize the
average absolute difference between the $\langle \pt \rangle$ values
at about a dozen selected time points from the new scheme and the
corresponding $\langle \pt \rangle$ values from the subdivision method
of Eq.~(\ref{subd5}),
where the selected time points are usually taken as the positions
of the symbols on the curves from the new scheme in
Figs.~\ref{fig:pt1}-\ref{fig:pt4}.

\begin{figure*}[htb]
\centerline{\includegraphics[scale=0.35]{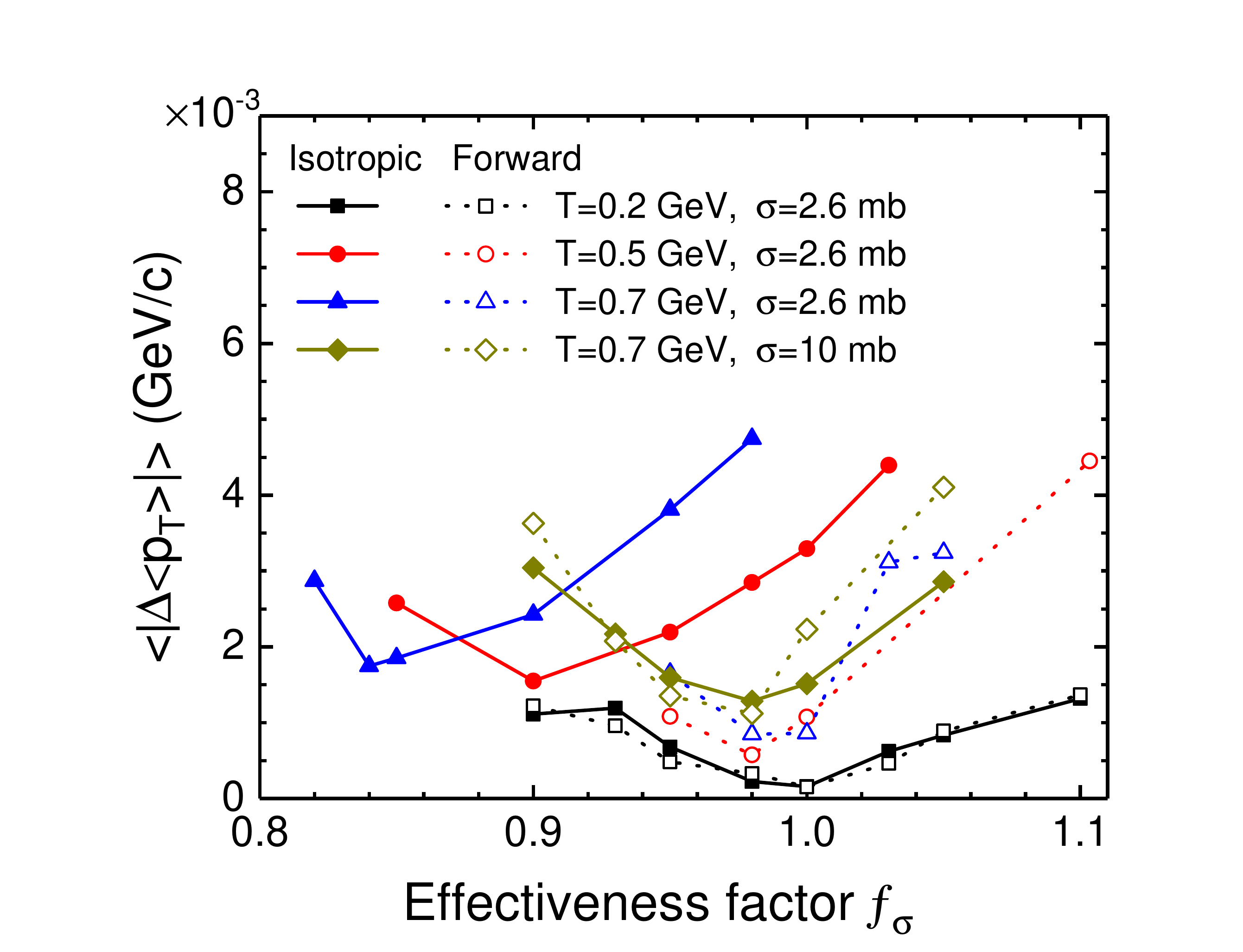}}
\caption{The average absolute difference in $\langle \pt \rangle$
versus $f_\sigma$, the effectiveness factor of cross section, for different cases.}
\label{fig:factor}
\end{figure*}

Figure~\ref{fig:factor} shows the average absolute difference in
$\langle \pt \rangle$ versus
the $f_\sigma$ value for several different cases,
where we can find the minimum position for each case.
The optimal $f_\sigma$ value, i.e., the value that gives the minimum $\langle \pt
\rangle$ difference, is given  for each case in Table~\ref{tablef}.
We can see that the optimal $f_\sigma$ value is 1.00 for the low
opacity case (at $T=0.2$ GeV and $\sigma=2.6$ mb), which is expected
because of the small causality violation.
On the other hand, the biggest deviation of the optimal $f_\sigma$
from unity occurs for the case of isotropic scatterings at
a moderate opacity (at $T=0.7$ GeV and $\sigma=2.6$ mb).
Note that this is also the case with the largest deviation of
the final ${\rm var}(\pt)/T^2$ value from the theoretical value, as
can be seen in Fig.~\ref{fig:chi}.
Also shown in Figs.~\ref{fig:pt2}-\ref{fig:pt4} are
the results from the subdivision method of Eq.~(\ref{subd5})
after using the corresponding optimal $f_\sigma$ values
(dot-dashed curves), which nicely match the time evolutions of
$\langle \pt \rangle$ from the new scheme.
For example, the optimal value $f_\sigma=0.93$ in
Fig.~\ref{fig:pt3} means that the new collision scheme for
forward-angle scatterings at $\sigma=10$ mb (and $T=0.7$ GeV) is
effectively equivalent to the ``exact'' parton subdivision method
that uses the cross section $\sigma=9.3$ mb (with the same scattering
angular distribution).

\begin{table}[h]
\caption{$f_\sigma$, the effectiveness factor of cross section, for
  different cases.}
\renewcommand\arraystretch{1.5}
\begin{tabular}{|c|c|c|c|c|}
\hline
T \& $\sigma$ values & 0.2 GeV \& 2.6 mb
 & 0.5 GeV \& 2.6 mb & 0.7 GeV  \& 2.6 mb  & 0.7 GeV \& 10 mb  \\
 & ($\chi=0.13$) &  ($\chi=2.0$)  &  ($\chi=5.5$) & ($\chi=41.$)\\
\hline
$f_\sigma$ for forward-angle scatterings & 1.00  & 0.98 & 0.98  & 0.98 \\
\hline
$f_\sigma$ for isotropic scatterings & 1.00    & 0.90   & 0.84 & 0.98\\
\hline
\end{tabular}
\label{tablef}
\end{table}

\subsection{Shear viscosity and the $\eta/s$ ratio}
\label{resultsB}

Transport coefficients such as the shear viscosity $\eta$ represent 
important properties of the created matter \cite{Schafer:2009dj}. It
is thus useful to evaluate the effect of our new collision scheme on
the shear viscosity $\eta$ and its ratio over the entropy density
$\eta/s$. The Green-Kubo relation~\cite{Green,Kubo} has been 
applied~\cite{Muronga:2003tb,Demir:2008tr,Fuini:2010xz,Wesp:2011yy,Li:2011xu}
to calculate the shear viscosity at or near equilibrium.  
Therefore we start with an equilibrium initial condition for shear viscosity
calculations in this section.
 
We calculate the shear viscosity according to the following form of
the Green-Kubo relation~\cite{Muronga:2003tb}:
\begin{equation}
\eta =\frac{V}{T} \int_{0}^{\infty}dt \,
\langle \bar{\pi }^{xy}(t+t^\prime)\,\bar{\pi }^{xy}(t^\prime) \rangle.
\label{gk}
\end{equation}
Here $\langle ... \rangle$ represents the time ($t^\prime$) and
ensemble average, and $\bar{\pi }^{xy}(t)$ represents the volume
averaged $xy$-component of the energy momentum tensor $\pi^{\mu \nu}$:
\begin{equation}
\pi^{\mu \nu}({\bm r},t)=\frac{1}{(2\pi)^3} 
\int d^{3}p \frac{p^{\mu}p^{\nu}}{p^{0}} f(\bm r,\bm p,t). 
\label{pimu}
\end{equation}
Since we do not consider parton potentials in the ZPC parton
cascade here, the volume average of $\pi^{xy}$ at a
given time $t$ can be written as  
\begin{equation}
\bar{\pi }^{xy}(t)=\frac{1}{V} \sum_{i=1}^{N}\frac{p_{i}^{x}p_{i}^{y}}{p^{0}_{i}},
\label{pit}
\end{equation}
where the sum is over all partons in the box at time $t$. 

It is known that the correlation function in Eq.~(\ref{gk}) damps
exponentially with time~\cite{Muronga:2003tb,Demir:2008tr,Wesp:2011yy}:
\begin{equation}
\langle \bar{\pi }^{xy}(t+t^\prime)\,\bar{\pi }^{xy}(t^\prime) \rangle
=\langle \bar{\pi }^{xy}(t^\prime)\,\bar{\pi }^{xy}(t^\prime) \rangle~e^{-t/\tau} 
\label{exptau}
\end{equation}
with $\tau$ being the corresponding relaxation time. Also, the 
average variance of $\bar{\pi }^{xy}$ in equilibrium is given by
\begin{equation}
\langle \bar{\pi }^{xy}(t^\prime)\,\bar{\pi }^{xy}(t^\prime) \rangle
=\frac{4}{15} \frac{\epsilon \,T}{V},  
\label{pi0pi0}
\end{equation}
where $\epsilon=3d_gT^4/\pi^2$ is the energy density of massless
gluons in equilibrium. 
We then have 
\begin{equation}
\eta =\frac{4}{15}\, \epsilon \,\tau.
\label{eta}
\end{equation}
So we extract the relaxation time $\tau$ from the
calculation of the correlation function in Eq.~(\ref{exptau}).
Specifically, the time and ensemble average of the correlation
function in our numerical calculations is obtained
as~\cite{Muronga:2003tb,Wesp:2011yy} 
\begin{equation}
\langle \bar{\pi }^{xy}(t+t^\prime)\,\bar{\pi }^{xy}(t^\prime) \rangle
=\left \langle \frac{1}{T_t} \int_{0}^{T_t}\bar{\pi
  }^{xy}(t+t^\prime)\,\bar{\pi }^{xy}(t^\prime) \, dt^\prime \right \rangle
\simeq \left  \langle
  \frac{1}{N_t} \sum_{j=0}^{N_t-1}\bar{\pi }^{xy}(i\Delta
  t+j\Delta t)\,\bar{\pi }^{xy}(j\Delta t) \right \rangle. 
\label{pitpi0}
\end{equation}
In the above, $N_t=T_t/\Delta t$, $t=i \Delta t$, 
and each of the last two $\langle ... \rangle$ symbols represents
the ensemble average. 
Here we typically choose $T_t \sim 30\tau, N_t \sim 200$, 
and extract $\tau$ from a fit to the normalized correlation function
over the the range $t \in [0, \sim \! 2\tau]$. 
Note that 
$\langle \bar{\pi }^{xy}(t+t^\prime)\,\bar{\pi }^{xy}(t^\prime) \rangle$ 
is often abbreviated as $\langle \bar{\pi }^{xy}(t)\,\bar{\pi }^{xy}(0)
\rangle$ in studies that use the Green-Kubo
relation~\cite{Muronga:2003tb,Demir:2008tr,Fuini:2010xz,Wesp:2011yy,Li:2011xu}.
In addition, for isotropic elastic collisions the shear viscosity and the
corresponding relaxation time of a massless
Maxwell$\textendash$Boltzmann gas in equilibrium can be calculated in
the Navier-Stokes approximation 
as~\cite{DeGroot:1980dk,Huovinen:2008te} 
\begin{equation}
\eta^{N\!S} \simeq 1.265\,\frac{T}{\sigma},
~\tau^{N\!S} \simeq \frac{1.582}{n\sigma}.
\label{etaNS}
\end{equation}

\begin{figure*}[htb]
\centerline{\includegraphics[scale=0.35]{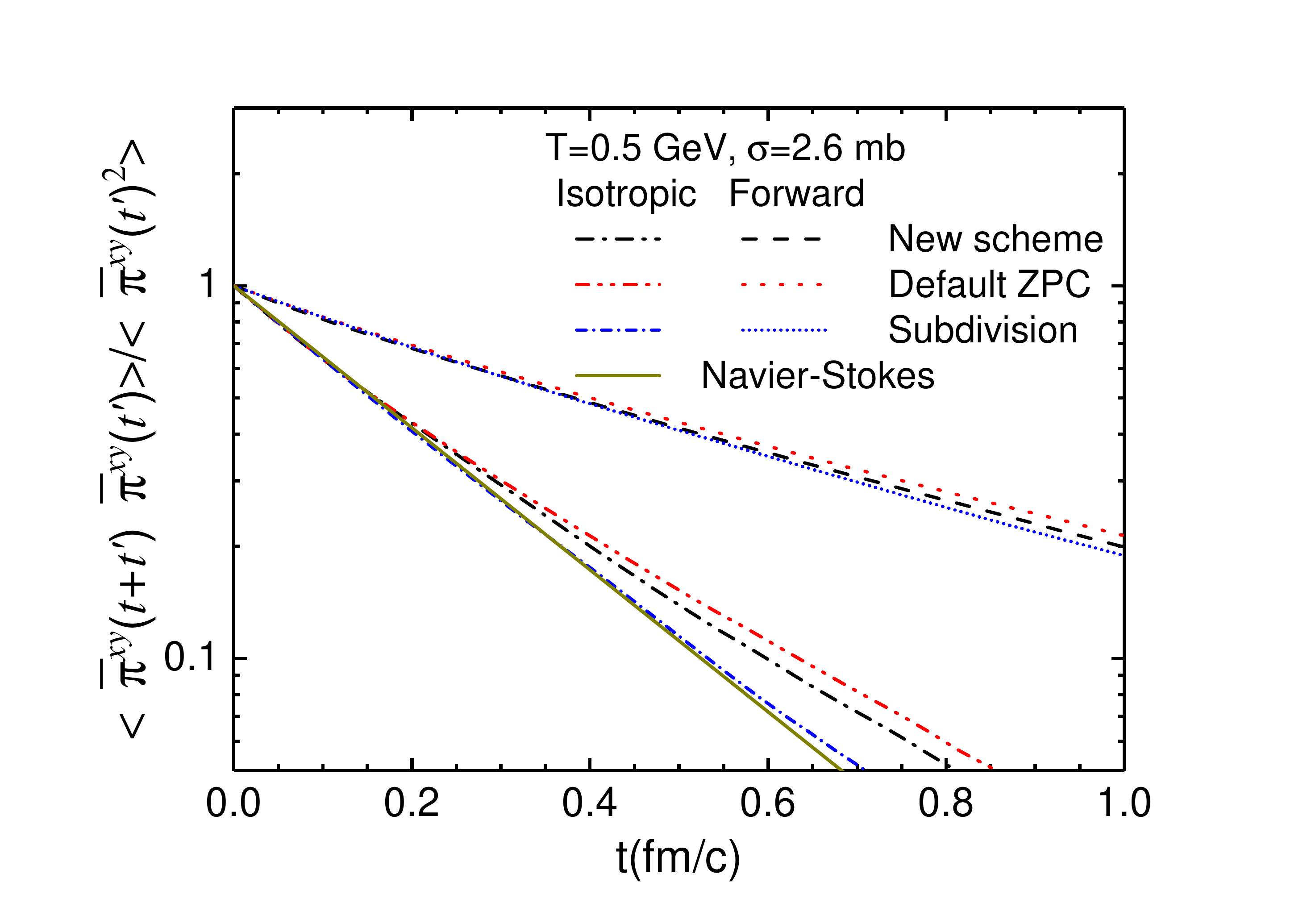}}
\caption{Normalized correlation functions from different collision
  schemes for isotropic scatterings  and forward-angle scatterings at 
$T=0.5$ GeV and $\sigma=2.6$ mb; the solid line represents the
Navier-Stokes expectation for isotropic scatterings.}
\label{fig:pit0nor}
\end{figure*}

We show in Fig.~\ref{fig:pit0nor} the normalized correlation functions
for the case of $T=0.5$ GeV and $\sigma=2.6$ mb, which corresponds to
opacity $\chi=2.0$. 
All the numerical results show the expected exponential damping with
time. For isotropic scatterings, we see that the result from the 
subdivision method is almost the same as the Navier-Stokes
expectation. On the other hand, the ZPC result without parton
subdivision using the new collision scheme or the default collision
scheme both damps a bit more slowly, which will lead to a bit larger
relaxation time and $\eta$ value than those from the parton
subdivision method. 
For forward-angle scatterings, Fig.~\ref{fig:pit0nor} also shows 
that the ZPC results without parton subdivision 
damp a bit more slowly than the result from parton subdivision. 
Furthermore, for both isotropic and forward-angle scatterings, 
the damping rates from the new scheme are closer to the subdivision
results than those from the default scheme. 

\begin{figure*}[htb]
\centerline{\includegraphics[scale=0.6]{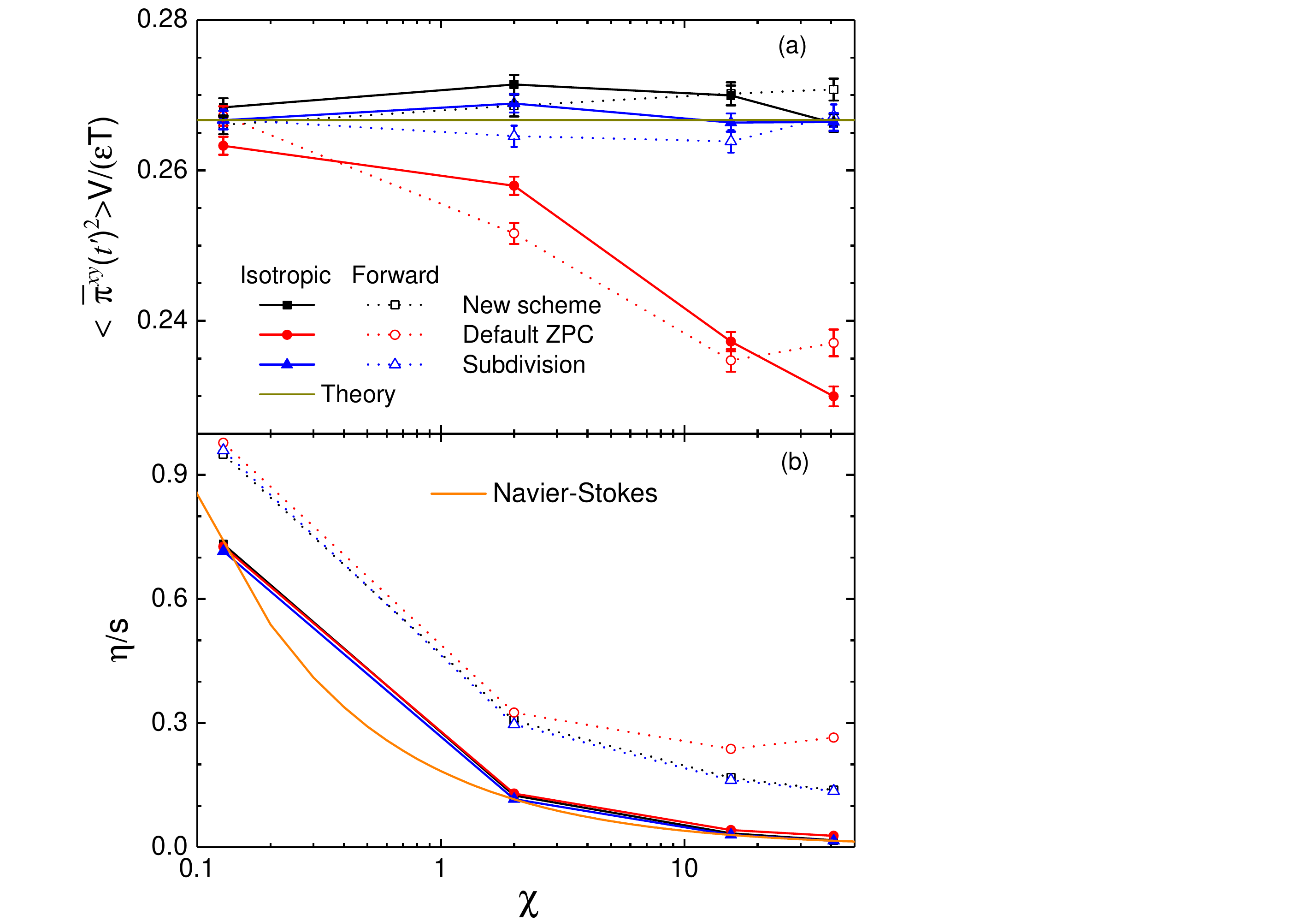}}
\caption{(a) Scaled average variance 
$\langle \bar{\pi }^{xy}(t^\prime)\,\bar{\pi }^{xy}(t^\prime) \rangle$
(with statistical error bars) versus opacity $\chi$ for different
cases in comparison with the theoretical value for equilibrium (solid
line). (b) The $\eta/s$ ratio for different cases 
versus opacity; the solid curve represents the Navier-Stokes 
expectation for isotropic scatterings.}
\label{fig:etas}
\end{figure*}

We also check our results of the average variance 
$\langle \bar{\pi }^{xy}(t^\prime)\,\bar{\pi }^{xy}(t^\prime)
\rangle$ in Fig.~\ref{fig:etas}(a) from different cases 
for both isotropic and forward-angle scatterings. The average variance
has been multiplied by the factor $V/(\epsilon\,T)$ because then its
theoretical value is 4/15 (solid line). 
The cases included in Fig.~\ref{fig:etas} are the same as those shown 
in Fig.~\ref{fig:chi} except for the omission of 
the case of $T=0.7$ GeV with $\sigma=2.6$ mb (at $\chi=5.5$). 
We see that  in general the results from the subdivision method agree
well with the theoretical expectation from low to very high opacities. 
The results from the new collision scheme also agree with the
theoretical value rather well. 
On the other hand, the average variance from the default ZPC collision
scheme deviates significantly from the theoretical value at finite
opacities (up to $\sim 16\%$ at the extreme opacity of $\chi=41$).  

Figure~\ref{fig:etas}(b) shows our $\eta/s$ results  
for these different cases as functions of opacity $\chi$. 
For isotropic scatterings of a massless Maxwell$\textendash$Boltzmann
gluon gas in equilibrium (where $s=4n$), we have the following 
Navier-Stokes expectation:
\begin{equation}
\left (\frac{\eta}{s} \right )^{N\!S} 
\simeq \frac{0.4633}{d_g^{1/3}\chi^{2/3}}
= \frac{0.1839}{\chi^{2/3}},
\label{etasNS}
\end{equation}
which only depends on the opacity $\chi$.
We see in Fig.~\ref{fig:etas}(b) that the subdivision results agree
well with the Navier-Stokes expectation (solid curve) for isotropic
scatterings, similar to the observation from Figs.~\ref{fig:pit0nor}
and \ref{fig:etas}(a). 
In addition, the results from the new collision scheme are very close
to the subdivision results for both forward-angle scatterings and
isotropic scatterings from small to large opacities. 
On the contrary, the extracted $\eta$ and $\eta/s$ values from the
default ZPC scheme can be significantly higher than the Navier-Stokes
expectation or the parton subdivision results at large opacities,
consistent with its lower collision rates at finite opacities as shown
in Fig.~\ref{fig:WT0}. 
However, the extracted $\eta$ value from the new scheme can also be 
somewhat higher than the Navier-Stokes expectation where the
corresponding collision rate is not lower than the theoretical value,
for example for the case shown in Fig.~\ref{fig:pit0nor}. 
Therefore in the presence of causality violation at finite opacities 
there are additional factors besides the collision rate that 
affect the shear viscosity of the parton system. 

\section{Conclusions}
\label{summary}

We have evaluated and then improved the accuracy of the ZPC
parton cascade for elastic scatterings inside a box. It is well known
that cascade solutions of the Boltzmann equation such as ZPC suffer
from the causality  violation at high densities and/or parton
scattering cross sections (i.e., large opacities), and that the parton
subdivision technique can be used to solve this problem. However,
parton subdivision alters event-by-event correlations and
fluctuations and is also computationally very expensive. In this
work we have found a collision scheme that is accurate enough without
parton subdivision and much better than the default ZPC collision
scheme. We first test a dozen different collision schemes
for the collision time(s) and ordering time of ZPC and find that the
default collision scheme does not accurately describe the equilibrium
momentum distribution at large opacities. We then find that a particular
collision scheme, the scheme that uses the minimum of the two
collision times as both the collision time and ordering time in the
global frame while calculating the closest approach distance in the
two-parton center of mass frame, can describe very accurately the
equilibrium momentum distribution as well as the time evolution
towards equilibrium, even at high opacities. 
In addition, we apply the Green-Kubo relation to calculate the shear
viscosity and the $\eta/s$ ratio for different cases, which also show
that the new collision scheme is more accurate than the default scheme
and agrees well with the theoretical expectation for isotropic
scatterings. Furthermore, we use a 
novel parton subdivision method to obtain the ``exact'' time evolution
of the momentum distribution towards equilibrium. This subdivision
method is valid for such box calculations, and it is so much more
efficient than the traditional subdivision method that we typically
use a subdivision factor of $10^6$. This work is the first
step towards the validation and improvement of the ZPC parton cascade
for scatterings in 3-dimensional expansion cases.

\begin{acknowledgments}
This work is supported in part by the National Natural Science
Foundation of China under Grants Nos. 11890714, 11835002, 11421505, 
11961131011 and the Key Research Program of the Chinese Academy of
Sciences under Grant No. XDB34030200  (GLM \& YGM), and the Chinese
Scholarship Council (XLZ). 
\end{acknowledgments}


\begin{thebibliography}{99}
\bibliographystyle{unsrt}

\bibitem{Adams:2005dq}
  J.~Adams {\it et al.} [STAR Collaboration],
  Nucl.\ Phys.\ A {\bf 757}, 102 (2005).

\bibitem{Adcox:2004mh}
  K.~Adcox {\it et al.} [PHENIX Collaboration],
  Nucl.\ Phys.\ A {\bf 757}, 184 (2005).

\bibitem{He:2015hfa}
  L.~He, T.~Edmonds, Z.~W.~Lin, F.~Liu, D.~Molnar and F.~Wang,
  Phys.\ Lett.\ B {\bf 753}, 506 (2016).

\bibitem{Lin:2015ucn}
  Z.~W.~Lin, L.~He, T.~Edmonds, F.~Liu, D.~Molnar and F.~Wang,
  Nucl.\ Phys.\ A {\bf 956}, 316 (2016).

\bibitem{Ko:2016ioz}
C.~M.~Ko and F.~Li,
Nucl. Sci. Tech. \textbf{27}, 140 (2016).

\bibitem{Jin:2018lbk}
X.~H.~Jin, J.~H.~Chen, Y.~G.~Ma, S.~Zhang, C.~J.~Zhang and C.~Zhong,
Nucl. Sci. Tech. \textbf{29}, 54 (2018).

\bibitem{Wang:2019vhg}
H.~Wang, J.~H.~Chen, Y.~G.~Ma and S.~Zhang,
Nucl. Sci. Tech. \textbf{30}, 185 (2019).

\bibitem{Geiger:1997pf}
  K.~Geiger,
  Comput.\ Phys.\ Commun.\  {\bf 104}, 70 (1997).

\bibitem{Gyulassy:1997zn}
  M.~Gyulassy, Y.~Pang and B.~Zhang,
  Prog.\ Theor.\ Phys.\ Suppl.\  {\bf 129}, 21 (1997).

\bibitem{Zhang:1997ej}
  B.~Zhang,
  Comput.\ Phys.\ Commun.\  {\bf 109}, 193 (1998).

 \bibitem{Molnar:2000jh}
  D.~Molnar and M.~Gyulassy,
  Phys.\ Rev.\ C {\bf 62}, 054907 (2000).

\bibitem{Xu:2004mz}
  Z.~Xu and C.~Greiner,
  Phys.\ Rev.\ C {\bf 71}, 064901 (2005).

\bibitem{Xu:2007aa}
  Z.~Xu and C.~Greiner,
  Phys.\ Rev.\ C {\bf 76}, 024911 (2007).

\bibitem{Lin:2004en}
  Z.~W.~Lin, C.~M.~Ko, B.~A.~Li, B.~Zhang and S.~Pal,
  Phys.\ Rev.\ C {\bf 72}, 064901 (2005).

\bibitem{Bzdak:2014dia}
  A.~Bzdak and G.~L.~Ma,
  Phys.\ Rev.\ Lett.\  {\bf 113}, 252301 (2014).

\bibitem{Li:2016ubw}
  H.~Li, L.~He, Z.~W.~Lin, D.~Molnar, F.~Wang and W.~Xie,
  Phys.\ Rev.\ C {\bf 96}, 014901 (2017).

\bibitem{Nagle:2017sjv}
  J.~L.~Nagle, R.~Belmont, K.~Hill, J.~O.~Koop, D.~V.~Perepelitsa, P.~Yin, Z.~W.~Lin and D.~McGlinchey,
  Phys.\ Rev.\ C {\bf 97}, 024909 (2018).

\bibitem{Kurkela3}
  A.~Kurkela, U.~A.~Wiedemann and B.~Wu,
  Phys.\ Lett.\ B {\bf 783}, 274 (2018);
  Eur.\ Phys.\ J.\ C {\bf 79}, 759 (2019);
  Eur.\ Phys.\ J.\ C {\bf 79}, 965 (2019).

\bibitem{Kodama:1983yk}
  T.~Kodama, S.~B.~Duarte, K.~C.~Chung, R.~Donangelo and R.~A.~M.~S.~Nazareth,
  Phys.\ Rev.\ C {\bf 29}, 2146 (1984).

\bibitem{Kortemeyer:1995di}
  G.~Kortemeyer, W.~Bauer, K.~Haglin, J.~Murray and S.~Pratt,
  Phys.\ Rev.\ C {\bf 52}, 2714 (1995).

\bibitem{Molnar:2019yam}
  D.~Molnar,
  arXiv:1906.12313 [nucl-th].

\bibitem{Lin:2014tya}
  Z.~W.~Lin,
  Phys.\ Rev.\ C {\bf 90}, 014904 (2014).

\bibitem{Zhang:1998tj}
  B.~Zhang, M.~Gyulassy and Y.~Pang,
  Phys.\ Rev.\ C {\bf 58}, 1175 (1998).

\bibitem{Cheng:2001dz}
  S.~Cheng, S.~Pratt, P.~Csizmadia, Y.~Nara, D.~Molnar, M.~Gyulassy, S.~E.~Vance and B.~Zhang,
  Phys.\ Rev.\ C {\bf 65}, 024901 (2002).

\bibitem{Zhang:1996gb}
  B.~Zhang and Y.~Pang,
  Phys.\ Rev.\ C {\bf 56}, 2185 (1997).

\bibitem{Wong:1982zzb}
  C.~Y.~Wong,
  Phys.\ Rev.\ C {\bf 25}, 1460 (1982).

\bibitem{Welke:1989dr}
  G.~Welke, R.~Malfliet, C.~Gregoire, M.~Prakash and E.~Suraud,
  Phys.\ Rev.\ C {\bf 40}, 2611 (1989).

\bibitem{Molnar:2004yh}
  D.~Molnar and P.~Huovinen,
  Phys.\ Rev.\ Lett.\  {\bf 94}, 012302 (2005).

\bibitem{Molnar:2001ux}
  D.~Molnar and M.~Gyulassy,
  Nucl.\ Phys.\ A {\bf 697}, 495 (2002);
  {\it ibid.} {\bf 703}, 893(E) (2002).

\bibitem{Pang:1997}
  Y.~Pang, https://karman.physics.purdue.edu/OSCAR-old/rhic/gcp/proc.html;
  General Cascade Program, in Proceedings of RHIC'96, CU-TP-815 (1997).

\bibitem{Alver:2010gr}
  B.~Alver and G.~Roland,
  Phys.\ Rev.\ C {\bf 81}, 054905 (2010);
  {\it ibid.} {\bf 82}, 039903(E) (2010).

\bibitem{Lipavsky:1986zz}
P.~Lipavsky, V.~Spicka and B.~Velicky,
Phys. Rev. B \textbf{34}, 6933-6942 (1986).

\bibitem{Semkat}
D. Semkat, D. Kremp, and M. Bonitz, 
J. Math. Phys. {\bf 41}, 7458 (2000).

\bibitem{Schafer:2009dj}
T.~Sch\"afer and D.~Teaney,
Rept. Prog. Phys. \textbf{72}, 126001 (2009).

\bibitem{Green}
M. S. Green, J. Chem. Phys. {\bf 22}, 398 (1954).

\bibitem{Kubo}
R. Kubo, J. Phys. Soc. Jpn. {\bf 12}, 570 (1957).

\bibitem{Muronga:2003tb}
A.~Muronga,
Phys. Rev. C  {\bf 69}, 044901 (2004).

\bibitem{Demir:2008tr}
N.~Demir and S.~A.~Bass,
Phys. Rev. Lett.  {\bf 102}, 172302 (2009).

\bibitem{Fuini:2010xz}
J.~Fuini, III, N.~S.~Demir, D.~K.~Srivastava and S.~A.~Bass,
J. Phys. G   {\bf 38}, 015004 (2011).

\bibitem{Wesp:2011yy}
C.~Wesp, A.~El, F.~Reining, Z.~Xu, I.~Bouras and C.~Greiner,
Phys. Rev. C  {\bf 84}, 054911 (2011).

\bibitem{Li:2011xu}
S.~Li, D.~Fang, Y.G.~Ma and C.~Zhou,
Phys. Rev. C  {\bf 84}, 024607 (2011).

\bibitem{DeGroot:1980dk}
S.~De Groot, W.~Van Leeuwen and C.~Van Weert,
{\it Relativistic Kinetic Theory: Principles and Applications} 
(North Holland, Amsterdam, 1980).

\bibitem{Huovinen:2008te}
P.~Huovinen and D.~Molnar,
Phys. Rev. C \textbf{79}, 014906 (2009).

\end{thebibliography}
\end{document}